\documentclass[aps,pra,twocolumn,showpacs,superscriptaddress]{revtex4-2}  % for review and submission

\usepackage{physics}
\usepackage{graphicx}  % needed for figures
\usepackage{dcolumn}   % needed for some tables
\usepackage{bm,amssymb,amsmath,dsfont}   % for math
\usepackage[colorlinks=true,breaklinks=true,allcolors=blue]{hyperref}
\usepackage{url}
\usepackage{txfonts}
\usepackage{bbold}

\usepackage{graphics}
\usepackage{pstricks}
\usepackage{tikz}
\usepackage{color}
\usepackage{xcolor}
\graphicspath{{Figures/}}

\usepackage{tabularx}
\usepackage{multirow}
\usepackage{array}
\newcolumntype{L}[1]{>{\raggedright\let\newline\\\arraybackslash\hspace{0pt}}m{#1}}
\newcolumntype{C}[1]{>{\centering\let\newline\\\arraybackslash\hspace{0pt}}m{#1}}
\newcolumntype{R}[1]{>{\raggedleft\let\newline\\\arraybackslash\hspace{0pt}}m{#1}}

\newcommand{\eq}[1]{(\ref{eq:#1})}
\newcommand{\Eq}[1]{Eq.\,\eq{#1}}

\newcommand{\Fig}[1]{Fig.~\ref{fig:#1}}
\newcommand{\fig}[1]{\ref{fig:#1}}

\newcommand{\Sect}[1]{Sect.~\ref{sec:#1}}

\newcommand{\Appendix}[1]{Appendix~\ref{app:#1}}

\newcommand{\App}[1]{App.~\ref{app:#1}}

\definecolor{applegreen}{rgb}{0.55, 0.71, 0.0}
\definecolor{byzantine}{rgb}{0.74, 0.2, 0.64}

\newcommand{\cred}[1]{{\color{red}{#1}}}
\newcommand{\cblu}[1]{{\color{blue}{#1}}}
\newcommand{\cgrn}[1]{{\color{applegreen}{}}}

\newcommand{\corg}[1]{{\color{orange}{}}}

\newcommand{\tbd}[1]{\cred{#1}}
\renewcommand{\tbd}[1]{}

\newcommand{\tbdso}[1]{\cblu{#1}}
\renewcommand{\tbdso}[1]{}

\newcommand{\IntermediateStep}[1]{&\cred{\textrm{\ (($===$ intermediate calc. steps $==>$))}}\nonumber\\ #1  
                                                           &\cred{\textrm{(($<=====================$))}}\nonumber\\}
\renewcommand{\IntermediateStep}[1]{}

\makeatletter
\let\cat@comma@active\@empty
\makeatother

\begin{document}

% the following line is for submission, including submission to the arXiv!!
%\hspace{5.2in} \mbox{Fermilab-Pub-04/xxx-E}

%==============================================================================
%==============================================================================
\title{Complex Langevin approach to interacting Bose gases}
\author{Philipp Heinen}
\affiliation{Kirchhoff-Institut f\"ur Physik,
             Ruprecht-Karls-Universit\"at Heidelberg,
             Im~Neuenheimer~Feld~227,
             69120~Heidelberg, Germany}

\author{Thomas Gasenzer}
\affiliation{Kirchhoff-Institut f\"ur Physik,
             Ruprecht-Karls-Universit\"at Heidelberg,
             Im~Neuenheimer~Feld~227,
             69120~Heidelberg, Germany}
\affiliation{Institut f\"{u}r Theoretische Physik,
		Universit\"{a}t Heidelberg, 
		Philosophenweg 16, 
		69120 Heidelberg, Germany}

%\date{\today}

\begin{abstract}
Quantitative numerical analyses of interacting dilute Bose-Einstein condensates are most often based on semiclassical approximations. 
Since the complex-valued field-theoretic action of the Bose gas does not offer itself to the direct application of standard Monte Carlo techniques, simulations beyond their scope by now almost exclusively rely on quantum-mechanical techniques.
Here we explore an alternative approach based on a Langevin-type sampling in an extended state space, known as complex Langevin (CL) algorithm.
While the use of the CL technique has a long-standing history in high-energy physics, in particular in the simulation of QCD at finite baryon density, applications to ultracold atoms are still in their infancy.
Here we examine the applicability of the CL approach for a one- and two-component, three-dimensional non-relativistic Bose gas in thermal equilibrium, below and above the Bose-Einstein phase transition.
By comparison with analytic descriptions at the Gaussian level, including Bogoliubov and Hartree-Fock theory, we find that the method allows computing reliably and efficiently observables in the regime of experimentally accessible parameters. 
Close to the transition, quantum corrections lead to a shift of the critical temperature which we reproduce within the numerical range known in the literature.
With this work, we aim to provide a first test of CL as a potential out-of-the-box tool for the simulation of experimentally realistic situations, including trapping geometries and multicomponent/-species models.
\end{abstract}

\pacs{%
%11.10.Wx 		%Finite-temperature field theory
%03.65.Db 	Functional analytical methods
%03.75.Kk, 	Dynamic properties of condensates; collective and hydrodynamic excitations, superfluid flow
%03.75.Lm 	  	%Tunneling, Josephson effect, Bose-Einstein condensates in periodic potentials, solitons, vortices, and topological excitations 
%05.60.Cd 	Classical transport
%05.70.Jk, 		%Critical point phenomena 
%25.75.-q, 	Relativistic heavy-ion collisions
%47.27.E-, 		%Turbulence simulation and modeling
%47.27.ef 	Field-theoretic formulations and renormalization
%47.27.T- 	Turbulent transport processes
%47.37.+q, 	Hydrodynamic aspects of superfluidity; quantum fluids
%67.85.De 		%Dynamic properties of condensates; excitations, and superfluid flow
%98.80.Cq, 	Particle-theory and field-theory models of the early Universe (including cosmic pancakes, cosmic strings, chaotic phenomena, inflationary universe, etc.)
}

\maketitle

%======================================================================================
%======================================================================================

\section{Introduction}
\label{sec:Introduction}
Interacting Bose-Einstein condensates are at the center of contemporary experimental as well as theoretical studies. 
Though much of the interest nowadays focuses on non-equilibrium properties, many open questions still remain for Bose gases in equilibrium. 
A tremendously successful description of the weakly interacting Bose gas is provided by Bogoliubov theory \cite{Pitaevskii2016bose}. 
However, there are many situations of interest where Bogoliubov theory is not very straightforward or even unfeasible to apply. 
These include  Bose gases in trapping potentials \cite{Giorgini1996a}, in lower dimensions \cite{Hutchinson2006phase}, with spin interactions \cite{Stamper-Kurn2013a.RevModPhys.85.1191} and at phase transitions, to name only a few. 

For such scenarios it would be convenient to have a method at hand to simulate the Bose gas from first principles. 
Unfortunately, the field theoretic path integral describing the Bose gas in equilibrium may not be simulated by straightforward Monte Carlo methods because the action is complex in general, an obstacle known as sign problem and plaguing numerical computations in a wide range of physical scenarios \cite{Henelius2000sign,Lombardo2007lattice,Alford2010mitigating,Aarts2009.PhysRevLett.102.131601}.

One possibility to circumvent this problem is to avoid the field theoretic (coherent state) path integral description of the non-relativistic Bose gas altogether. 
This approach is pursued by the path integral Monte Carlo (PIMC) technique \cite{Pilati2006equation,Spada2021.PhysRevA.105.013325}, which treats the problem of interacting bosons in a quantum-mechanical rather that a field-theoretic framework. 
This method provides exact simulations also for strongly interacting bosons, while its computational cost increases polynomially with the number of particles to be captured. 

Other methods stick to the field-theoretic formulation, where there is no limitation on particle number, but employ a semi-classical approximation.
These include the truncated Wigner \cite{Sinatra2002truncated,Blakie2008a} as well as the projected Gross-Pitaevskii equation (PGPE) algorithm \cite{Davis2001simulations,Davis2002b}.

On the other hand, there are numerical methods available that promise to directly tackle the sign problem in the evaluation of path integrals. These include, inter alia, the diagrammatic Monte Carlo method \cite{boninsegni2006worm,vanhoucke2010diagrammatic}, the density of states algorithm \cite{Fodor2007the,Gattringer2016approaches}, the dual variables approach \cite{Endres2007method,Gattringer2016approaches} and the fixed node Monte Carlo method \cite{Vanbemmel1994fixed,Foulkes2001quantum,Rothstein2019introduction}.
Two particularly promising, completely general and model-independent approaches are the Lefschetz thimbles \cite{Cristoforetti2012new,Pawlowski2021simulating} as well as the complex Langevin (CL) algorithm \cite{Parisi1983complex,Klauder1985spectrum,Seiler2018status}. Both involve the artificial complexification of real fields, thereby doubling the number of degrees of freedom of the original problem, somewhat similar in spirit to the positive P-representation approach \cite{Gilchrist1997positive,Deuar2006a,Deuar2006b}.

In this work we will focus on the complex Langevin algorithm. 
The method has a long-standing history in the simulation of lattice QCD at finite chemical potential \cite{Aarts2008stochastic,Sexty2014simulating,Kogut2019applying,Ito2020complex}. 
More recently, it has also found applications in condensed matter physics, mainly in the simulation of non-relativistic fermions \cite{Loheac2017third,Rammelmuller2018finite}. 
In contrast, works applying CL in the context of bosonic ultracold atoms are scarce, with the notable exception of \cite{Hayata2015complex,Attanasio2020thermodynamics}.

By performing a systematic study of the three-dimensional interacting Bose gas we here want to demonstrate that CL can also be a useful tool for simulations of ultracold bosonic atoms, in particular because it provides a possibility to perform \textit{exact} simulations within the field-theoretic description of interacting bosons. 
This paves the way for evaluating observables in equilibrium Bose-Einstein condensates of arbitrary particle number from first principles, without the need of employing any approximation scheme.

%======================================================================================
%======================================================================================
\section{The complex Langevin method}
The path integral
\begin{align}
	\label{eq:Z}
	Z=\int \mathcal{D}\phi\,\exp(-S[\phi])
	\,,
\end{align}
for a theory described by the action $S[\phi]$ for the variable or field $\phi$, generalizes the partition sum in statistical mechanics and thus offers itself for the numerical evaluation of observables according to
\begin{align}
	\label{eq:ExpValue}
	\langle\mathcal{O}\rangle_{S}
	=Z^{-1}\int \mathcal{D}\phi\,\exp(-S[\phi])\,\mathcal{O}(\phi)
	\,.
\end{align}
In contrast to classical statistical mechanics, where the measure involves a positive-definite probability distribution, the action $S$ in the quantum path integrals \eq{Z} and \eq{ExpValue} is in general, however, not real-valued. 
This renders a straightforward Monte Carlo evaluation of the expectation value \eq{ExpValue} unfeasible, an obstacle known as the sign problem. 
Many, by order of magnitude large, but positive and negative contributions can contribute to the sum, both in the real and imaginary parts, to give an eventually comparatively small result.
Evaluating this sum generically is exponentially hard \cite{Berger2021complex}.

A particular approach developed for tackling this difficulty is the complex Langevin (CL) method \cite{Berger2021complex}. 
It exploits the well-known equivalence between path integrals and the stochastic Langevin equation as well as the fact that the latter can be formulated also for complex actions. 
The Langevin approach is well established in statistical mechanics, corresponding to the case of a real-valued action $S[\phi]$ describing, in the simplest case, the dynamics of a real-valued field $\phi$. 
Expectation values of the form \eq{ExpValue} can be computed by evolving the field $\phi$ along a fictitious time $\vartheta$ according to the stochastic Langevin equation
\begin{align}
	\label{eq:rl_eq}
	\frac{\partial\phi}{\partial\vartheta}
	=-\frac{\delta S}{\delta\phi}+\eta(\vartheta)
	\,,
\end{align}
with $\eta(\vartheta)$ being a Wiener noise averaging to $\langle\eta(\vartheta)\rangle=0$ and subject to the Markovian covariance $\langle \eta(\vartheta)\eta(\vartheta')\rangle=2\delta(\vartheta-\vartheta')$, and by taking the average of $\mathcal{O}(\phi(\vartheta))$ along the direction of $\vartheta$. 
\Eq{rl_eq} can be discretized as 
\begin{align}
	\phi_{i+1}
	=\phi_i-\Delta\vartheta\,\frac{\delta S}{\delta\phi}+\sqrt{2\Delta\vartheta}\,\eta_i
	\,
\end{align}
with the $\eta_i$ being Gaussian random variables with zero mean and standard deviation $1$ and $\Delta\vartheta$ being the discretization of the Langevin time.

The CL method extends this scheme for the case of a complex action $S\in\mathbb{C}$.
Observing that \Eq{rl_eq} leads into the complex plane even for the case of a real-valued field $\phi$, one complexifies each field component, $\phi\to\phi_R+i\phi_I$, which are evolved according to twice as many Langevin equations as before,
\begin{align}
	\label{eq:cl_eq}
	\frac{\partial\phi_R}{\partial\vartheta}
	&=-\Re\left[\frac{\delta S}{\delta\phi}\right]+\eta_R(\vartheta)
	\,,\\
	\frac{\partial\phi_I}{\partial\vartheta}
	&=-\Im\left[\frac{\delta S}{\delta\phi}\right]+\eta_I(\vartheta)
	\,,
\end{align}
with the noise being subject to
\begin{align}
	\langle\eta_R(\vartheta)\rangle
	&=\langle\eta_I(\vartheta)\rangle=0
	\,,\nonumber\\ 
	\langle\eta_R(\vartheta)\eta_R(\vartheta')\rangle
	&=2N_R\delta(\vartheta-\vartheta')
	\,,\qquad
	N_R-N_I=1
	\,,
	\\ 
	\langle\eta_I(\vartheta)\eta_I(\vartheta')\rangle
	&=2N_I\,\delta(\vartheta-\vartheta')
	\,. \nonumber
\end{align}
Hence, the CL equations involve, in general, a stochastic force in both the real and imaginary directions. 
The two noise contributions must fulfill the condition $N_R-N_I=1$ in order ensure that the path integral involves a single integration ``direction" as in a line integral \cite{Kades:2021hir}. 
It has been shown, however, that it is numerically most convenient to set $N_R=1$ and $N_I=0$ \cite{Aarts2010complex.PhysRevD.81.054508}.

While it is a straightforward exercise to demonstrate the equivalence of the Langevin approach to the original path integral for real actions, its validity in the case of complex actions cannot be established rigorously.
It is known to fail in certain cases, either because of the occurrence of runaway trajectories in the complex plane or because the process converges to an unphysical result \cite{Aarts2010complex.PhysRevD.81.054508,Hayata2016complex,Nishimura2015new}, which both represent characteristic consequences of the sign problem.
Several methods have been developed to at least ameliorate the problem of runaways \cite{Aarts:2010vk,Seiler2013gauge,Attanasio2019dynamical,Alvestad2021stable}.
Furthermore, there is substantial numerical evidence that the method can indeed give reliable results in a broad range of physical settings \cite{Berger2021complex}. 
While the sign problem eventually traces back to the unevadable complexity of Hilbert space, there have been several attempts to settle the issue by defining criteria for the correctness of the method \cite{Gausterer1993mechanism,Aarts2010complex.PhysRevD.81.054508,Nagata2016argument,Salcedo2016does,Scherzer2019complex,Scherzer2020controlling}.

In this paper we rather adopt a hands-on approach and test the complex Langevin approach at the exemplary system of a dilute Bose gas in three dimensions above and below the condensation phase transition. 
Since the upper critical dimension of the dilute Bose gas is $d_\mathrm{up}=2$, the three-dimensional system is expected to be well-described by mean-field theory close to the phase transition, and thus also further away from it.
Hence we evaluate momentum distributions and dispersion relations at different chemical potentials and compare them to analytical predictions from quantum field theory at the mean-field level, using the Bogoliubov and Hartree-Fock approximations below and above the transition, respectively.
We moreover explore the more intricate properties near the phase transition and reproduce the shift of the critical temperature within the limits reported in the literature.  

In this way we can demonstrate that the CL method represents a promising and easy-to-implement technique for computing many-body observables for interacting bosonic quantum gases in thermal equilibrium. 
The method can be straightforwardly extended to trapped configurations, multiple components and more intricate interaction terms, and in principle also lower dimensions where quantum correlations are expected to play a stronger role and are expected to provide a more stringent test bed of the method. 

Since CL is not the only method in the context of ultracold bosonic atoms that employs Langevin-type field equations, some remarks about other stochastic approaches to Bose-Einstein condensates are in order. 
These mainly include methods based on the positive-P representation, which were already mentioned above, as well as the stochastic Gross-Pitaevskii equation (SGPE) \cite{Stoof2001dynamics,Gardiner2002the,Gardiner2003the,Cockburn2009stochastic,Gautam2014finite}. 
What distinguishes CL from both of them is that it attempts to directly compute the Feynman path integral. 
The stochastic dynamics thus takes place in a fictitious, unphysical time, during which the Langevin process explores the space of field configurations, i.e., samples from the possible values of the quantum field at each point in space and physical time (the physical time can be either imaginary as in thermal equilibrium or real as in non-equilibrium). 
In contrast, both the positive-P method and the SGPE are formulated in the physical time of the system, and the stochastic dynamics can be regarded as an actual physical evolution. 
A further distinctive feature of CL which it shares with the positive-P method but not with the SGPE approach is that it attempts to transform the computationally hard problem of sampling from a non-positive-definite distribution (as it arises generally when one wants to perform full quantum simulations) to sampling from a positive-definite distribution in a phase space with twice as many degrees of freedom. 
Thus, for a complex Bose field $\psi$, both CL and the positive-P method require to evolve \textit{four} real fields instead of two. 
In contrast, the SGPE does not feature any artificial extension of the number of degrees of freedom. 
The stochastic term here arises from the coupling of the condensate to the cloud of thermally excited particles. 
By virtue of including this interaction with the thermal background, the SGPE goes beyond a pure mean-field description, but it does not attempt to solve the full quantum problem from first principles and can rather be classified as a semi-classical method \cite{Gardiner2003the}. 
In this regard it differs from CL and the positive-P approach, which both \textit{in principle} provide exact simulations of the full quantum problem but are not applicable in all situations of interest due to the intrinsic and inevitable complexity of quantum physics.

%======================================================================================
%======================================================================================
\section{Complex-Langevin simulations of a thermal dilute Bose gas}
%
%======================================================================================
\subsection{The dilute $\mathcal{N}$-component Bose gas}
We study a non-relativistic $\mathcal{N}$-component Bose gas with U$(\mathcal{N})$ symmetric interactions in thermal equilibrium at inverse temperature $\beta=1/T$ and chemical potential $\mu$ \footnote{
Throughout the paper we use natural units where $\hbar=k_\mathrm{B}=1$}. 
The action, expressed in terms of the complex Bose fields $\psi_a(\tau,\mathbf{x})$, $a=1,\dots,\mathcal{N}$, defined on the $4$-dimensional Euclidean manifold of imaginary time $\tau$ and position $\mathbf{x}$, reads 
\begin{align}
	\label{eq:action}
	S[\psi_a,\psi^*_a]
	=\int \limits_0^\beta \mathrm{d}\tau\int \mathrm{d}^3x \left[\psi_a^*\partial_\tau\psi_a+\mathcal{H}(\psi_a,\psi^*_a)\right]
	\,,
\end{align}
with Hamiltonian density
\begin{align}
	\label{eq:Hamiltonian}
	\mathcal{H}(\psi_a,\psi^*_a)
	=\frac{1}{2m}\nabla\psi_a^*\cdot\nabla\psi_a-\mu\,\psi^*_a\psi_a+\frac{g}{2}(\psi^*_a\psi_a)^2
	\,,
\end{align}
where summation over the field components, $a=1,\dots,\mathcal{N}$, is implied, and $m$ is the mass of the particles.
For a dilute atomic gas in $d=3$ dimensions, the dimensionful coupling $g=4\pi a/m$ is commonly defined in terms of the $s$-wave scattering length $a$, while the relevant measure for the coupling strength is the dimensionless gas parameter 
\begin{align}
	\label{eq:diluteness}
	\eta\equiv\sqrt{\rho a^3}
	\,.
\end{align}
Being defined in terms of the average density of particles $\rho$, it relates $a$ to the mean interparticle spacing $\rho^{-1/3}$ and evaluates, in typical experimental settings, to a number on the order of $\eta\sim 10^{-3}$. 
Hence, the model is weakly interacting and expected to be well described by low-order perturbative approximations, at least substantially below and above the condensation phase transition.

%======================================================================================
\subsection{Observables}
We are eventually interested in computing correlation functions of operators $\mathcal{O}$ as 
\begin{align}
	\label{eq:pi}
	&\left\langle\mathcal{O}_{1}\cdots\mathcal{O}_{n}\right\rangle
	\nonumber\\
	&\ =\frac{\int \mathcal{D}\psi_a \mathcal{D}\psi_a^*\,\mathrm{e}^{-S[\psi_a,\psi^*_a]}
	\,\mathcal{O}_{1}(\psi_a,\psi^*_a)\cdots\mathcal{O}_{n}(\psi_a,\psi^*_a)}
	{\int \mathcal{D}\psi_a \mathcal{D}\psi_a^*\,\exp(-S[\psi_a,\psi^*_a])}
	\,.
\end{align}
Since the action \eq{action} is, in general, not purely real-valued due to the "Berry phase" term $\psi_a^*\partial_\tau\psi_a$,  a straightforward Monte Carlo evaluation of the path integral \eq{pi} is unfeasible, which thus forms an exemplary problem we plan to tackle using the complex Langevin approach. 

In the following, we will focus on uniform dilute Bose gases in cubic spatial volumes subject to periodic boundary conditions, such that we can evaluate observables most conveniently in momentum space.
The main object we consider is the occupation number of the mode with wave number $\mathbf{k}$,
\begin{align}
	\label{eq:occnum}
	f(\mathbf{k})=\beta^{-1}\int_{0}^{\beta}\mathrm{d}\tau\,\langle \psi^*_a(\mathbf{k},\tau)\psi_a(\mathbf{k},\tau)\rangle\,,
\end{align}
where summation over $a=1,\dots,\mathcal{N}$ is implied.
Exploiting the homogeneity in $\tau$ direction, we average the $\tau$-local expectation value over imaginary time in order to gain a higher statistical accuracy.

Close to the transition and in the condensate phase, $f$ is expected to be much larger than one, $f(\mathbf{k})\gg1$, for all modes in the semi-classical regime of momenta below the temperature scale $\mathbf{k}^{2}/2m\ll T$, where also classical statistical methods such as the truncated Wigner approximation \cite{Sinatra2002truncated,Polkovnikov2010a}, or the PGPE algorithm \cite{Davis2001simulations} are applicable. 
At larger momenta, however, where occupancies are small, $f(\mathbf{k})\ll1$, the quasi-classical approximation is expected to fail, and one needs either genuine quantum statistical methods or single-particle techniques as the path-integral Monte Carlo (PIMC) algorithm \cite{Pilati2006equation}.
The PIMC approach is formulated in a quantum-mechanical rather than a quantum field-theoretic framework and  thus is limited to comparatively small total particle numbers.
We will demonstrate that the CL method is applicable in both the strongly and the weakly occupied regimes.

For the single-component system, $\mathcal{N}=1$, we additionally extract the dispersion by evaluating 
\begin{align}
	\label{eq:disp}
	\omega(\mathbf{k})
	=\beta^{-1}\int_{0}^{\beta}\mathrm{d}\tau\,
	\sqrt{-\frac{\partial_\tau\partial_{\tau'}\langle\psi^*(\mathbf{k},\tau)\psi(\mathbf{k},\tau')\rangle|_{\tau'=\tau}}
	{\langle\psi^*(\mathbf{k},\tau)\psi(\mathbf{k},\tau')\rangle|_{\tau'=\tau}}}
	\,,
\end{align}
The dispersion will give a first measure of the energy of elementary excitations of the system.
Also here, exploiting the homogeneity in $\tau$ direction, we average $\omega(\mathbf{k})$ over $\tau$ in order to gain a higher statistical accuracy.

%======================================================================================
\subsection{\label{sec:CLImplementation}Implementation of the CL simulations}
All our simulations of a one- ($\mathcal{N}=1$) and a two-component ($\mathcal{N}=2$) Bose gas were performed on an $N_xN_yN_zN_{\tau}=64^3\times N_{\tau}$ space-time lattice, with varying discretization $a_\tau=\beta/N_{\tau}$ along the imaginary-time direction, ranging between $N_\tau=16$ and $N_\tau=64$.
Details on the discretization of the action \eq{action} and the derivation and form of the corresponding Langevin equations are provided in \Appendix{langeq}. 
For simplicity, we fix the temporal units by choosing $(2m)^{-1}=a_\mathrm{s}$ in terms of the inverse of the lattice spacing $a_\mathrm{s}$. 
The temperature and thus $\beta$ are then given in units of $a_\mathrm{s}$, fixing, for a given $N_{\tau}$ also the imaginary-time lattice length $a_{\tau}$ in units of $a_\mathrm{s}$.
The Langevin time step is set to $\Delta\vartheta=10^{-3} a_\mathrm{s}$. 

In each run, we prepared the system in some initial state, as described in the subsequent sections. 
For example, above the transition, we chose the vacuum state, setting $\psi(\tau,\mathbf{x})=0$ at each point on the spatial and temporal lattice.
We then propagated the discretized Langevin equations, details of which are provided in \App{langeq}. After a certain equilibration time that varies with the physical scenario, we can compute averages of the observables along the field trajectories.

The computational cost of a single Langevin time step scales as the total space-time lattice size.
In addition to this, also the Langevin evolution time required to reach a fixed statistical precision scales as $N_{\tau}$, such that the total computational cost scales as $\sim N_{x}N_{y}N_{z}N_\tau^2$. 
The latter can be inferred from the fact that the action, discretized in $\tau$ direction, is of the form 
\begin{align}
	S=\sum_{i=1}^{N_{\tau}}\int \mathrm{d}^3x\left\{\psi^*_{i+1}(\psi_{i+1}-\psi_i)+a_{\tau}\, H(\psi^*_{i+1},\psi_i)\right\}\,,
\end{align}
where $a_{\tau}=\beta/N_\tau$ is the imaginary-time step size and $H$ the discretized Hamiltonian, cf.~\App{langeq} for details. 
Increasing $N_\tau$ for fixed $\beta$ implies a correspondingly smaller drift term and thus a slower exploration of the space of field configurations, such that a longer Langevin evolution time is required for obtaining the same statistical accuracy.

In order to speed up the simulation, we parallelized the computation of the Langevin drift and the propagation of the field according to the Langevin equations on GPUs, in a way such that every GPU thread updates a single lattice point. Working on NVIDIA V100 cards, it takes about $40$ hours of wall time to evolve a one-component $64^3\times16$ lattice for $10^7$ time steps and to extract the momentum spectrum after each time step. While for most scenarios we evolved for $\sim10^7$ steps, we increased this number to $10^8$ steps for the simulations close to the transition in order to gain a higher statistical accuracy needed for the precise determination of the transition point.

Statistical errors are generally estimated from the variance of several independent runs.
For further details on the numerical extraction of observables, see \App{spectrumanddispersion}.

%======================================================================================
\subsection{\label{sec:free}Ideal gas -- dependence on time discretization}
Before we move on to evaluating the observables defined above for an interacting dilute Bose gas at and away from the condensation transition, we perform, as a first benchmark of our approach, CL simulations of an ideal Bose gas. 
We use this, in particular, to discuss the errors induced by the discretization of imaginary time, corresponding to a truncation of the Fourier representations beyond a highest  Matsubara frequency. 

For the ideal gas, the coupling vanishes, $g=0$, and we tune the chemical potential to $\mu=-0.1\,a_\mathrm{s}^{-1}$, above the phase transition, where the condensate fraction vanishes.
We choose the temperature to be $T=1.25\,a_\mathrm{s}^{-1}$, which, together with $\mu$ corresponds to a total density $\rho=0.054\,a_s^{-3}$ and thus a critical temperature $T_\mathrm{c}=[\rho/\zeta(3/2)]^{2/3}2\pi/m=0.95\,a_\mathrm{s}^{-1}$ and a thermal de Broglie wave length of $\lambda_T= \sqrt{2\pi/mT} = 3.17 \,a_\mathrm{s}$. 

We simulate for a Langevin time of $\vartheta_\mathrm{max}=10^6\,a_\mathrm{s}$, and begin the averaging after an equilibration time $\vartheta_0=10^5\,a_\mathrm{s}$. Additionally, we average over 10 independent runs.
In each run, we prepare the system in the vacuum state, setting $\psi(\tau,\mathbf{x})=0$ at each point on the lattice. 

%==============================================================
\begin{figure}
	\includegraphics[width=0.95\columnwidth]{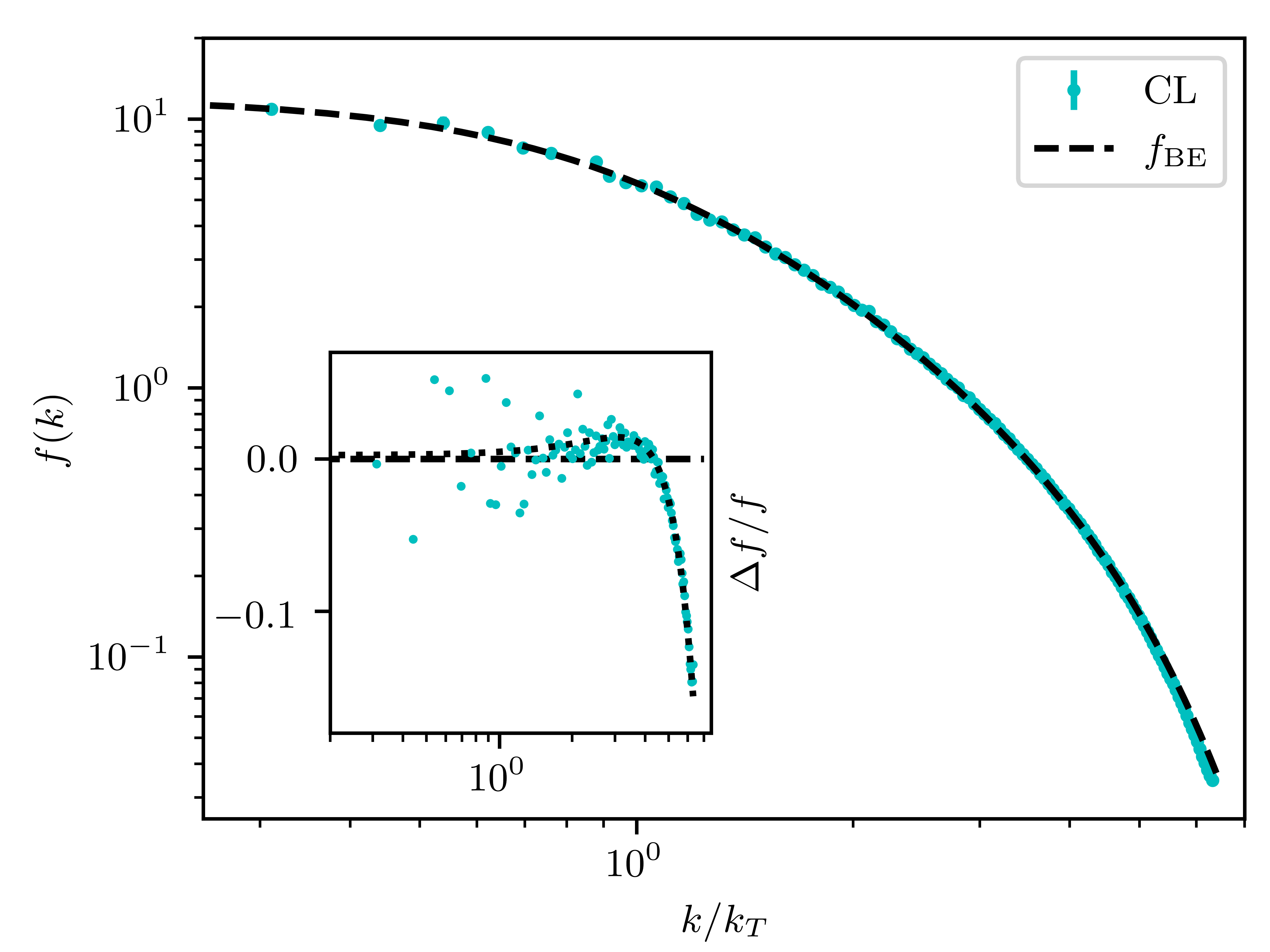}
	\caption{Angle-averaged momentum spectrum $f_\mathrm{CL}(k)$ of a single-component, non-interacting gas above the Bose-Einstein phase transition, at a temperature $T=1.25\,a_\text{s}^{-1}$ and chemical potential $\mu=-0.1\,a_\text{s}^{-1}$ (in units of the lattice constant $a_\text{s})$, as obtained by means of the complex Langevin method on a $64^{3}\times N_{\tau}$  lattice, with $N_\tau=16$ imaginary-time points.
	The momentum is measured in units of the inverse thermal de Broglie wave length $k_T=\lambda_T^{-1}=\sqrt{mT/2\pi}=0.315\,a_\mathrm{s}^{-1}$. 
	The black dashed line represents the Bose-Einstein distribution $f_\mathrm{BE}(k)$. 
	The inset shows the relative deviation $(\Delta f/f)^\mathrm{CL}_\mathrm{BE}=(f_\mathrm{CL}-f_\mathrm{BE})/f_\mathrm{BE}$ of the numerical and continuum analytic distributions. Note that the statistical error is generally higher for smaller $k$ because the corresponding momentum shells contain less modes to average over.
	The dotted black line in the inset represents the relative deviation as predicted from the analytical finite-$N_\tau$ computation.
	Note that here (and in all plots above the transition) for better visibility the spectrum is shown only up to $k_\mathrm{max}=2\,a_\mathrm{s}^{-1}$, i.e. excluding the "corners" of the momentum lattice. A version including the corners, i.e. up to $k_\mathrm{max}=2\sqrt{3}\,a_\mathrm{s}^{-1}$, is shown in \Fig{freespectrumwcorners}, \App{details}.
	}
	\label{fig:freespectrum}
\end{figure}
%==============================================================
The resulting, angular averaged  momentum distribution $f_\mathrm{CL}(k)=f_\mathrm{CL}(|\mathbf{k}|)=\langle f_\mathrm{CL}(\mathbf{k})\rangle_{\Omega_\mathbf{k}}$ for $N_\tau=16$ is shown in \Fig{freespectrum}. 
At large, the results for the distribution agree excellently with the analytic expression, 
\begin{align}
	\label{eq:freeBED}
	f_\mathrm{BE}(\mathbf{k})
	&=\frac{1}{\mathrm{e}^{\,\beta[\varepsilon(\mathbf{k})-\mu]}-1}
	\,,
\end{align}
with free single-particle dispersion
\begin{align}
	\varepsilon(\mathbf{k})=\frac{\mathbf{k}^2}{2m}
	\,,
\end{align}
over the whole range of momenta, while a weak deviation is seen at the smallest occupancies in the Boltzmann tail.
We amplify this discrepancy by showing, in the inset, the relative deviation $(\Delta f/f)^\mathrm{CL}_\mathrm{BE}=(f_\mathrm{CL}-f_\mathrm{BE})/f_\mathrm{BE}$ on the same logarithmic momentum scale. 

As we recall from \App{Matsubara}, the occupation number of the ideal gas, on a discrete imaginary-time lattice, can be written as 
\begin{align}
	\label{eq:fFiniteNtau}
	f_\mathrm{BE}(\mathbf{k};N_{\tau})
	&=\sum_{n=0}^{N_{\tau}-1}\frac{1}{{N_\tau}(\mathrm{e}^{2\pi\mathrm{i} n/{N_\tau}}-1)+\beta(\varepsilon(\mathbf{k})-\mu)}
	\,.
\end{align}
We show the relative deviation $(\Delta f/f)^{N_{\tau}}_\mathrm{BE}=(f_\mathrm{BE}(\mathbf{k})-f_\mathrm{BE}(\mathbf{k};N_{\tau}))/f_\mathrm{BE}(\mathbf{k})$ as a dotted line in the inset of \Fig{freespectrum}, which agrees on average with the deviation $(\Delta f/f)^\mathrm{CL}_\mathrm{BE}$ between the CL data and the exact result. 
This, in particular, demonstrates that the systematic increase of the relative numerical error in the low-occupancy region at large momenta can be attributed to the truncation of the Matsubara series.

\begin{figure}	
	\includegraphics[width=0.95\columnwidth]{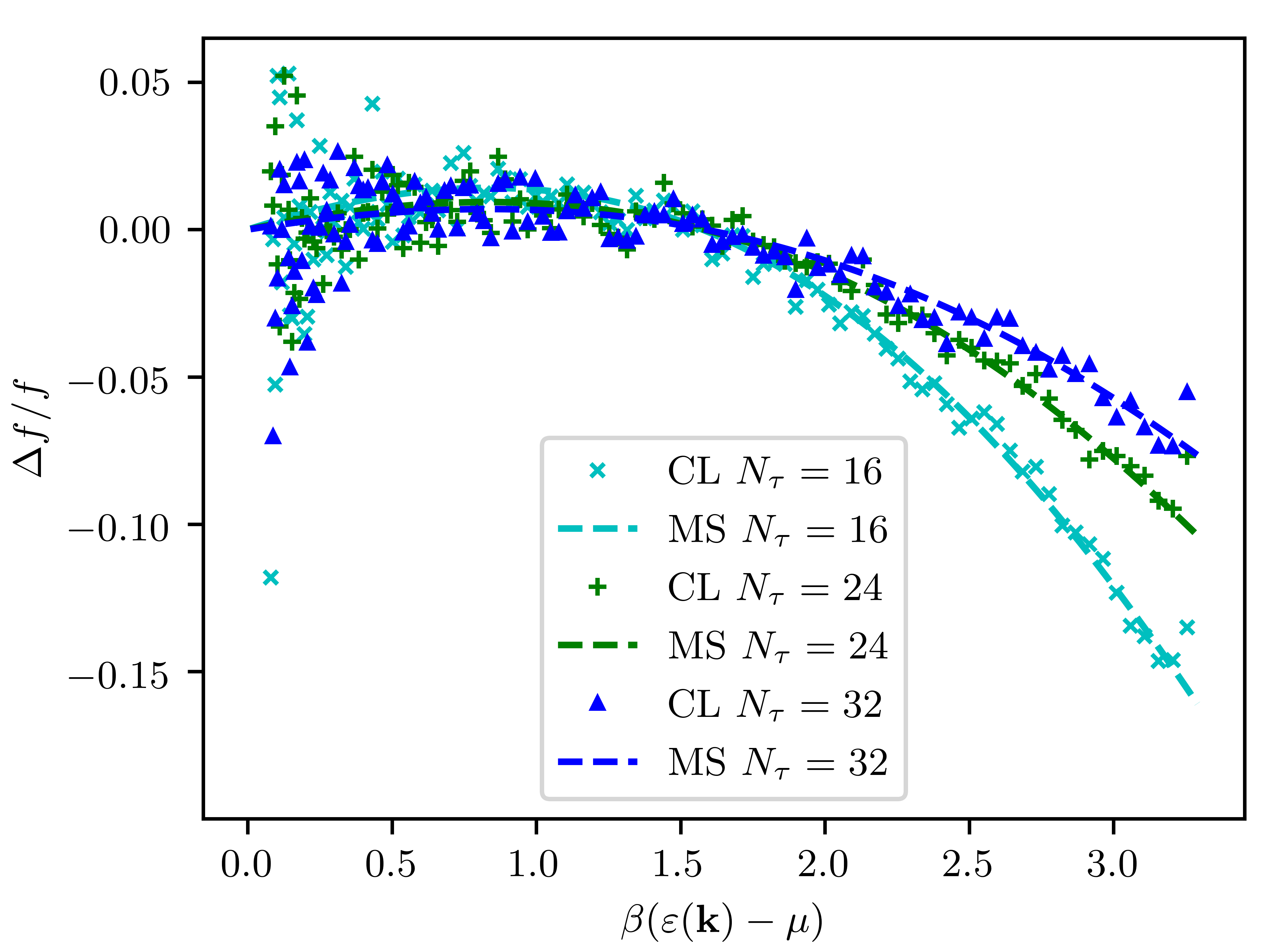}
	\caption{Comparison between the relative deviation $(\Delta f/f)^\mathrm{CL}_\mathrm{BE}=(f_\mathrm{CL}-f_\mathrm{BE})/f_\mathrm{BE}$ of the numerical and analytic, \eq{freeBED}, Bose-Einstein distributions (CL, points) and the finite-size deviation $(\Delta f/f)^{N_{\tau}}_\mathrm{BE}=[f_\mathrm{BE}(\mathbf{k};N_{\tau})-f_\mathrm{BE}(\mathbf{k})]/f_\mathrm{BE}(\mathbf{k})$ due to the truncation of the Matsubara series (MS, dashed lines), as a function of $\beta(\varepsilon(\mathbf{k})-\mu)$.
	The data is obtained for the same parameters as in \Fig{freespectrum}, but for three different $N_{\tau}\in\{16,24,32\}$.
	}
	\label{fig:comparison}
\end{figure}

In order to further corroborate this observation, we have performed runs with a different number of lattice points $N_{\tau}$ along the imaginary-time direction. 
The respective deviations $(\Delta f/f)^\mathrm{CL}_\mathrm{BE}$ and $(\Delta f/f)^{N_{\tau}}_\mathrm{BE}$ for $N_\tau=16$, $24$, $32$ are compared with each other in \Fig{comparison}. 
As one can see, the deviation of the CL data from the exact Bose-Einstein distributions decreases with increasing $N_\tau$, as predicted by the analytical finite-$N_\tau$ results shown as dashed lines.
%==============================================================

%==============================================================

\begin{figure}
	\includegraphics[width=0.95\columnwidth]{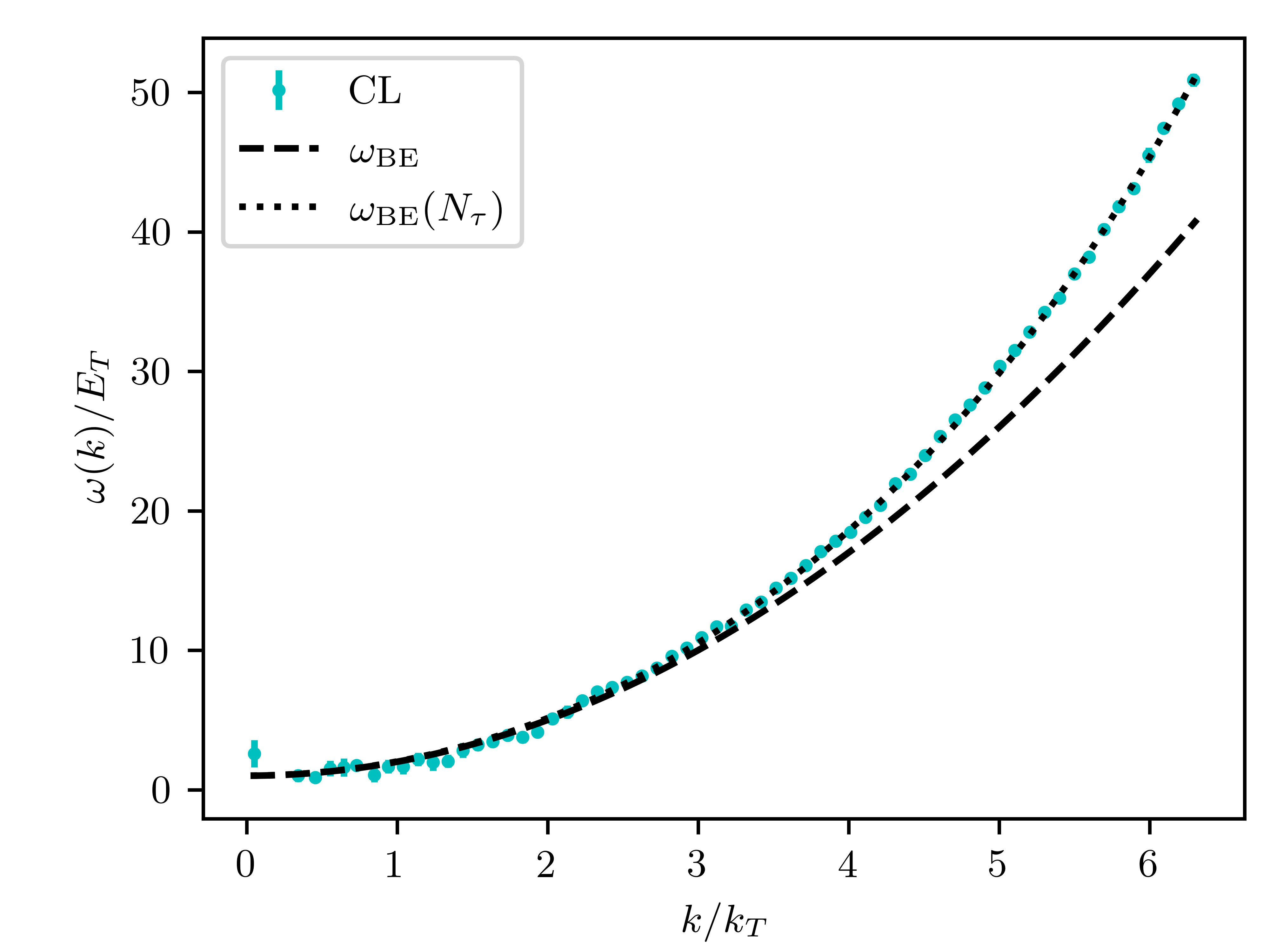}
	\caption{Dispersion $\omega_\mathrm{CL}(k)$ of the ideal Bose gas, as obtained according to \Eq{disp} from the same data as used in \Fig{freespectrum} and averaged over the angular orientations of $\mathbf{k}$, in units of $E_T\equiv k_T^2/2m$. Note that for better visibility we doubled the width of the momentum bins in comparison to \Fig{freespectrum}.
	The black dashed line indicates the free-gas dispersion \eq{freegasdisp}, whereas the corresponding prediction \eq{freegasdispNtau} on a discrete temporal lattice is shown as a dotted black line.
	The momentum is measured in units of the inverse thermal de Broglie wave length $k_T=0.315\,a_\mathrm{s}^{-1}$.
	} 	
	\label{fig:freedisp}
\end{figure}

\begin{figure}	
	\includegraphics[width=0.95\columnwidth]{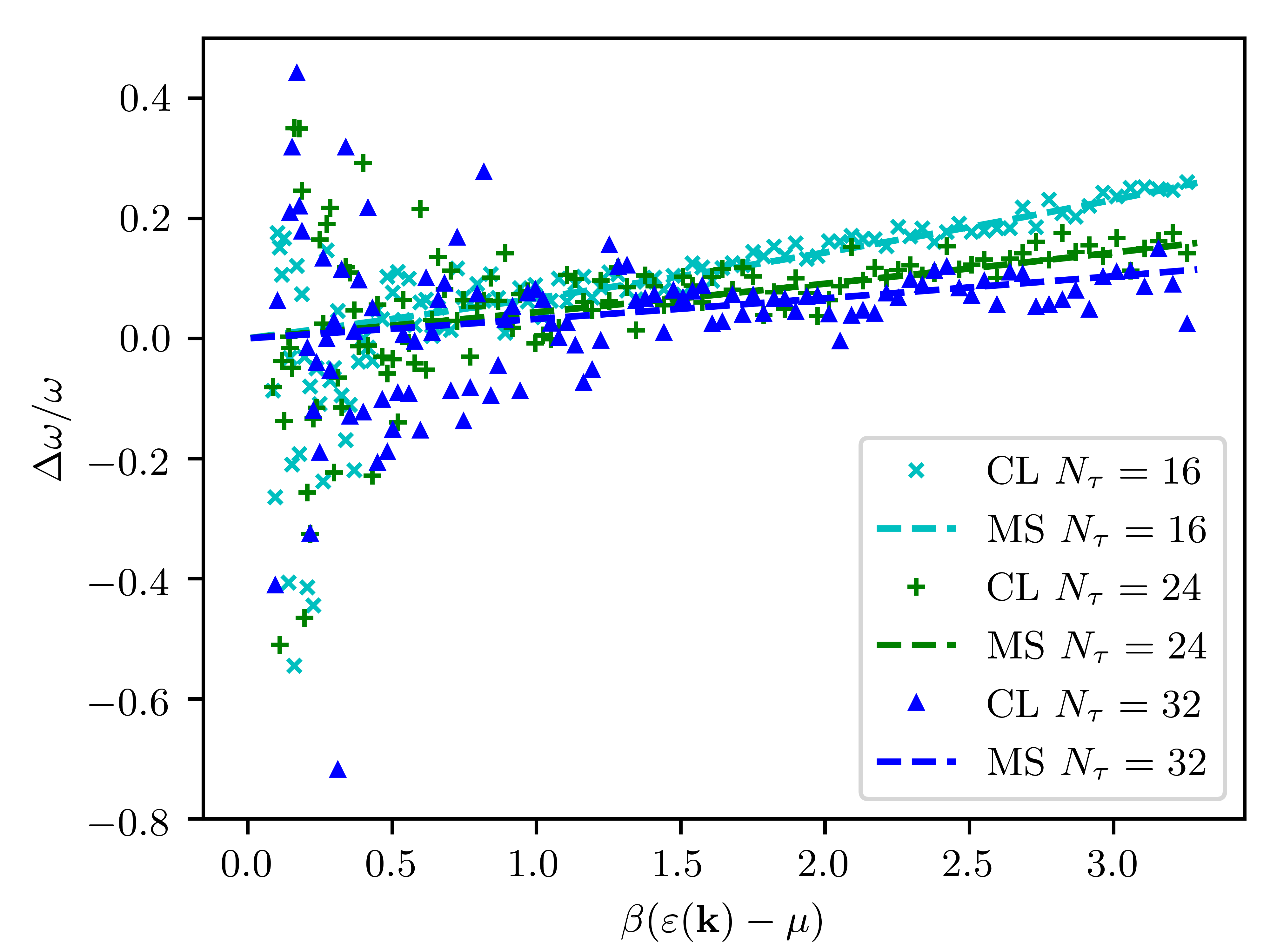}
	\caption{Comparison between the relative deviation $(\Delta \omega/\omega)^\mathrm{CL}_\mathrm{BE}=(\omega_\mathrm{CL}-\omega_\mathrm{BE})/\omega_\mathrm{BE}$ of the numerical and analytic, \eq{freegasdisp}, free gas dispersion and the finite-size deviation $(\Delta \omega/\omega)^{N_{\tau}}_\mathrm{BE}=[\omega_\mathrm{BE}(\mathbf{k};N_{\tau})-\omega_\mathrm{BE}(\mathbf{k})]/\omega_\mathrm{BE}(\mathbf{k})$ due to the truncation of the Matsubara series, as a function of $\beta(\varepsilon(\mathbf{k})-\mu)$.
	The data is obtained for the same parameters as in \Fig{freespectrum}, but for three different $N_{\tau}\in\{16,24,32\}$.
	}
	\label{fig:comparison2}
\end{figure}

The dispersion of the ideal Bose gas, $\omega_\mathrm{CL}({k})=\omega_\mathrm{CL}(|\mathbf{k}|)=\langle \omega_\mathrm{CL}(\mathbf{k})\rangle_{\Omega_\mathbf{k}}$, obtained as an angular average over \eq{disp},  is shown for $N_\tau=16$ in \Fig{freedisp}.
The continuum dispersion is given by the basic free-gas kinetic energy, shifted by the chemical potential, 
\begin{align}
	\label{eq:freegasdisp}
	\omega_\mathrm{BE}(\mathbf{k})
	&= \varepsilon(\mathbf{k})-\mu
	\,,
\end{align}
while the finite-size dispersions result, as described in \App{Matsubara}, as
\begin{align}
	\label{eq:freegasdispNtau}
	\omega_\mathrm{BE}(\mathbf{k};N_\tau)
	=\sqrt{\frac{-N_\tau^2T^{2}}{f(\mathbf{k};N_\tau)}
	\sum_{n=0}^{N_{\tau}-1}\frac{2\mathrm{e}^{-2\pi \mathrm{i} n/{N_\tau}}-\mathrm{e}^{-4\pi \mathrm{i} n/{N_\tau}}-1}
	{{N_\tau}(\mathrm{e}^{2\pi \mathrm{i} n/{N_\tau}}-1)+\beta[\varepsilon(\mathbf{k})-\mu]}}
	\,.
\end{align}
%
%==============================================================
%==============================================================
The deviation between the CL and continuum analytic results is visibly larger, but can be attributed, as for the momentum spectrum $f$, to the finite lattice resolution in imaginary-time direction and thus to the corresponding truncation of the sum over the Matsubara frequencies. This is again corroborated by a comparison of the deviation for different $N_\tau$, as shown in \Fig{comparison2}.

%==============================================================
%==============================================================

%======================================================================================
\subsection{Interacting gas above the BEC phase transition}
\label{sec:IntBECabovePT}
Having performed first benchmark computations for an ideal Bose gas, we now turn to a weakly interacting, i.e., dilute gas and use the complex Langevin approach to calculate momentum spectra and dispersions for chemical potentials and temperatures away from and near the condensation phase transition.
We start with considering the case above the transition, where we can compare our numerical results to analytic predictions on the basis of the Hartree-Fock (HF) Gaussian approximation. 
Subsequently, we benchmark our CL results to Bogoliubov theory below the transition, for both one- and two-component systems.
Finally, we will explore the vicinity of the phase transition, where we can obtain a benchmark beyond mean-field theory, by computing the relative shift of the critical temperature due to interactions, a quantity that is sensitive to non-Gaussian fluctuations.

%==============================================================
\begin{figure}
	\includegraphics[width=0.95\columnwidth]{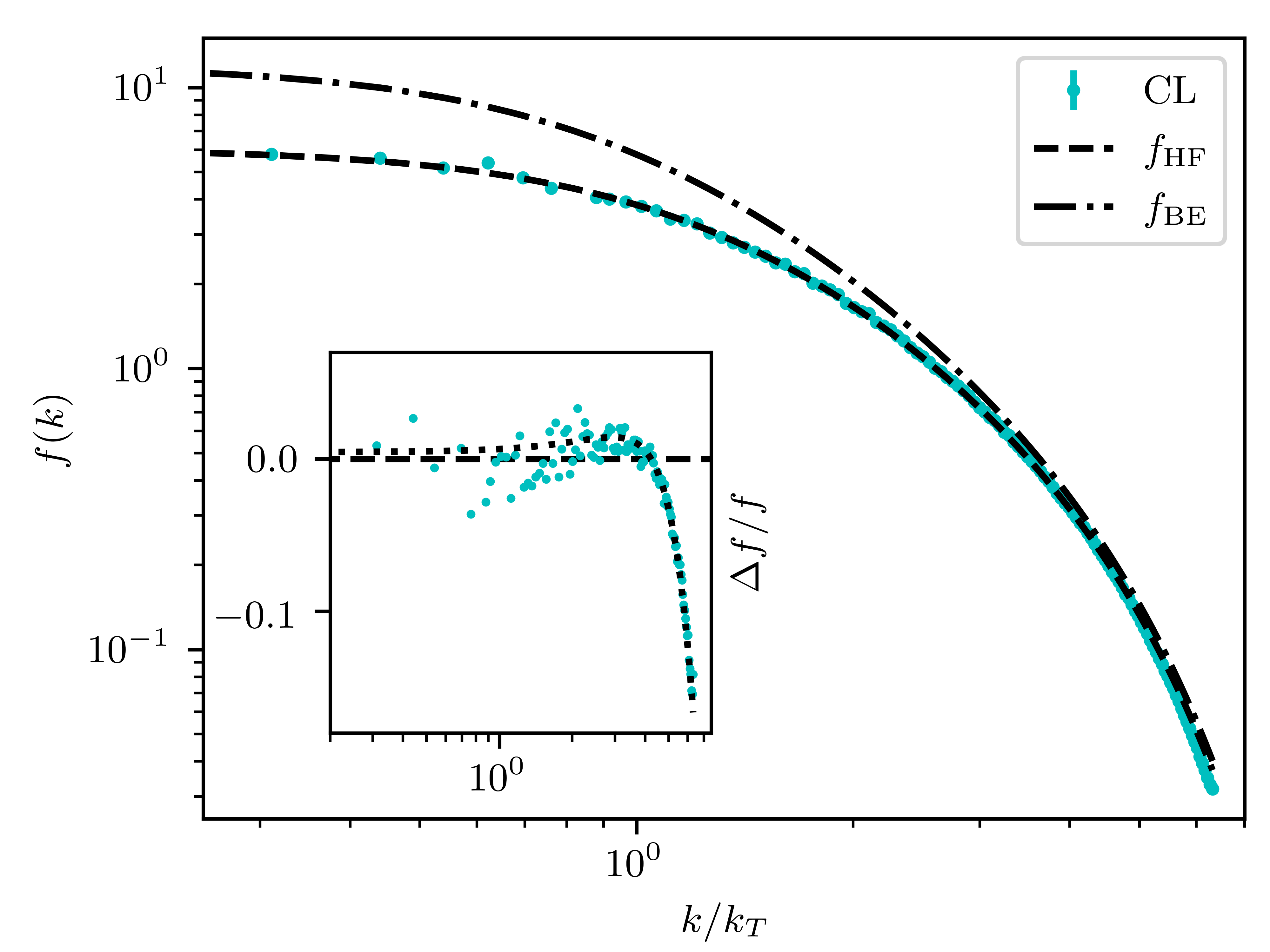}
	\caption{Angle-averaged momentum spectrum $f_\mathrm{CL}(k)$ of an interacting single-component gas above the BEC phase transition. 
	Temperature and chemical potential are chosen as in \Fig{freespectrum}, the coupling strength is $g=1.0\,a_\mathrm{s}^2$, giving rise to a diluteness $\eta\approx 1.7\cdot 10^{-3}$.
	The momentum is measured in units of the inverse thermal de Broglie wave length $k_T=\lambda_T^{-1}=\sqrt{mT/2\pi}=0.315\,a_\mathrm{s}^{-1}$. 
	The CL data is shown on a $64^{3}\times N_{\tau}$ lattice, with $N_{\tau}=16$ and averaged over $10$ runs.
	The spectrum fits well with the one calculated within the Hartree-Fock approximation (dashed black line). 
	For comparison we include the Bose-Einstein distribution of the corresponding non-interacting system (dash-dotted black line). 
	The inset shows the relative deviation $(\Delta f/f)^\mathrm{CL}_{\mathrm{HF}}=[f_\mathrm{CL}({k})-f_\mathrm{HF}({k})]/f_\mathrm{HF}({k})$. 
	The dotted black line in the inset represents the deviation $(\Delta f/f)^{N_{\tau}}_{\mathrm{HF}}=[f_\mathrm{HF}({k};N_{\tau})-f_\mathrm{HF}({k})]/f_\mathrm{HF}({k})$ of the finite-$N_{\tau}$ from the $N_{\tau}\to\infty$ version of the HF spectrum. 
	}
	\label{fig:abovetrans}
\end{figure}
%==============================================================
Above the critical point, we use the CL approach to compute the observables for the same choice of temperature and chemical potential, but with a non-vanishing, positive coupling constant $g=1\,a_\mathrm{s}^2$, quantifying the repulsive contact interactions between the bosons. 
Our simulations return, for the chosen parameters, a density of $\rho= (0.0447\pm0.002)\,a_\mathrm{s}^{-3}$, where we have extracted $\rho$ as described in \App{details}.
The corresponding diluteness parameter $\eta$, \Eq{diluteness}, is found to be  $\eta\approx 1.7\cdot 10^{-3}$, which is within the range typically reached in cold-gas experiments.
Note, that, with the basic implementation of the method as described here, in the grand-canonical ensemble formulation, we can freely choose temperature and chemical potential, while the mean energy and density are obtained from the numerical data.
This procedure can be straightforwardly extended to systematically determine the Lagrange parameters from given mean thermodynamic quantities.

\Fig{abovetrans} shows the momentum spectrum of the dilute gas in analogy to the ideal-gas spectrum depicted in \Fig{freespectrum}, on a $64^{3}\times N_{\tau}$ lattice, with $N_{\tau}=16$.
The main panels demonstrates good agreement with results obtained within the Hartree-Fock (HF) approximation \cite{Dalfovo1999a}, as described in the following. 

Expanding the  Hamiltonian \eq{Hamiltonian} in leading order of a mean-field approximation corresponds to replacing the quartic term by
\begin{align}
	\label{eq:HF}
	(\psi^{*}\psi)^2
	\to 4\langle \psi^{*}\psi\rangle \psi^{*}\psi+\left(\langle \psi^{*}\psi^{*}\rangle \psi\psi+\mathrm{c.c.}\right)
	\,.
\end{align}
Above the transition, we can neglect the in general complex anomalous densities $\langle\psi\psi\rangle$ such that the resulting Hamiltonian describes, again, a non-interacting Bose gas, albeit with a shifted chemical potential
\begin{align}
	\label{eq:mueffHF}
	\mu'=\mu-2g\langle \psi^{*}\psi\rangle\,.
\end{align}
The mean particle density $\rho=\langle |\psi|^2\rangle$, which equals the momentum integral over the Bose-Einstein distribution, hence must be determined self-consistently, that is, by solving the equation
\begin{align}
	\label{eq:selfkons}
	\rho=\int\frac{\mathrm{d}^{3}k}{(2\pi)^{3}}\,f_\mathrm{BE}(\mathbf{k};\mu')
	=\lambda_{T}^{-3}\,g_{3/2}\left(\mathrm{e}^{\,\beta(\mu-2g\rho)}\right)
	\,.
\end{align}
To determine $\rho$ and from this $\mu'$, we numerically solve \Eq{selfkons}, evaluating the polylogarithm directly.
We obtain $\rho_\mathrm{HF}=0.0454\,a_\mathrm{s}^{-3}$ which can be compared to the density obtained from our CL simulations as indicated above. 
The density corresponds to $\mu'=-0.191\,a_\mathrm{s}^{-1}$, which gives the ideal-gas Bose-Einstein distribution $f_\mathrm{HF}(k)$ shown as the dashed line in \Fig{abovetrans}, which significantly deviates from the dash-dotted distribution $f_\mathrm{BE}$ for $\mu$. 

The inset of \Fig{abovetrans} shows the relative deviation $(\Delta f/f)^\mathrm{CL}_{\mathrm{HF}}=[f_\mathrm{CL}({k})-f_\mathrm{HF}({k})]/f_\mathrm{HF}({k})$ from the expected Hartree-Fock result and compares it to the deviation $(\Delta f/f)^{N_{\tau}}_{\mathrm{HF}}=[f_\mathrm{HF}({k};N_{\tau})-f_\mathrm{HF}({k})]/f_\mathrm{HF}({k})$ of the finite-$N_{\tau}$ from the $N_{\tau}\to\infty$ version of the HF spectrum, on a temporal lattice of $N_{\tau}=16$ points. 
This demonstrates good agreement over the whole range of momenta.

%======================================================================================
\subsection{Bose gas in the condensed phase}
Before we move on to studying the region close to the BE phase transition, we repeat the previous analysis away from criticality, in the condensed phase.
In a condensate, the pattern of possible collective excitations fundamentally changes due to spontaneous symmetry breaking.
While the Hugenholtz-Pines (or Goldstone) theorem implies the elementary excitations to be gapless in the zero-energy limit, the interactions suppress density vs. phase fluctuations, such that long-range phase coherence prevails which renders the long-wave-length excitations sound like.
In the case $\mathcal{N}>1$ of more than one internal component of the U$(\mathcal{N})$ symmetric model \eq{Hamiltonian}, however, additional Goldstone excitations are possible, which correspond to relative density variations between the components.
These excitations are not subject to the suppression by interactions, which depend on the total density $\sum_{a=1}^{\mathcal{N}}\psi_{a}^{*}\psi_{a}$ only, and are therefore similar to the motion of non-interacting bosons.
To explore our algorithm for this case, too, we consider, in the following,  systems with both $\mathcal{N}=1$ and $\mathcal{N}=2$.

%======================================================================================
\subsubsection{One-component system}
We start again with the single-component system, $\mathcal{N}=1$, for which we now choose a temperature $T=0.625\,a_\mathrm{s}^{-1}$, chemical potential $\mu=0.5\,a_\mathrm{s}^{-1}$, and coupling constant $g=0.1\,a_\mathrm{s}^2$. 
This results in a total density of $\rho\approx\mu/g=5\,a_\mathrm{s}^{-3}$, corresponding to a diluteness $\eta = 0.56\cdot 10^{-3}$.
As before, we work on a $64^{3}\times N_{\tau}$  lattice, with different choices of the imaginary-time size $N_\tau=16$.

It turned out, though, that initializing the field as $\psi(\tau,\mathbf{x};\vartheta=0)=0$ for the Langevin evolution, as we did for all runs above the phase transition, the Langevin process is not able to build up a condensate, i.e., a highly over-occupied $\mathbf{k}=0$ mode even at temperatures far below the transition, at least not within the limited simulation time available.
This could be expected for large systems as the relative size of the phase space covered by the condensate mode is inversely proportional to the system size.
%==============================================================

\begin{figure}[th!]
	\includegraphics[width=0.95\columnwidth]{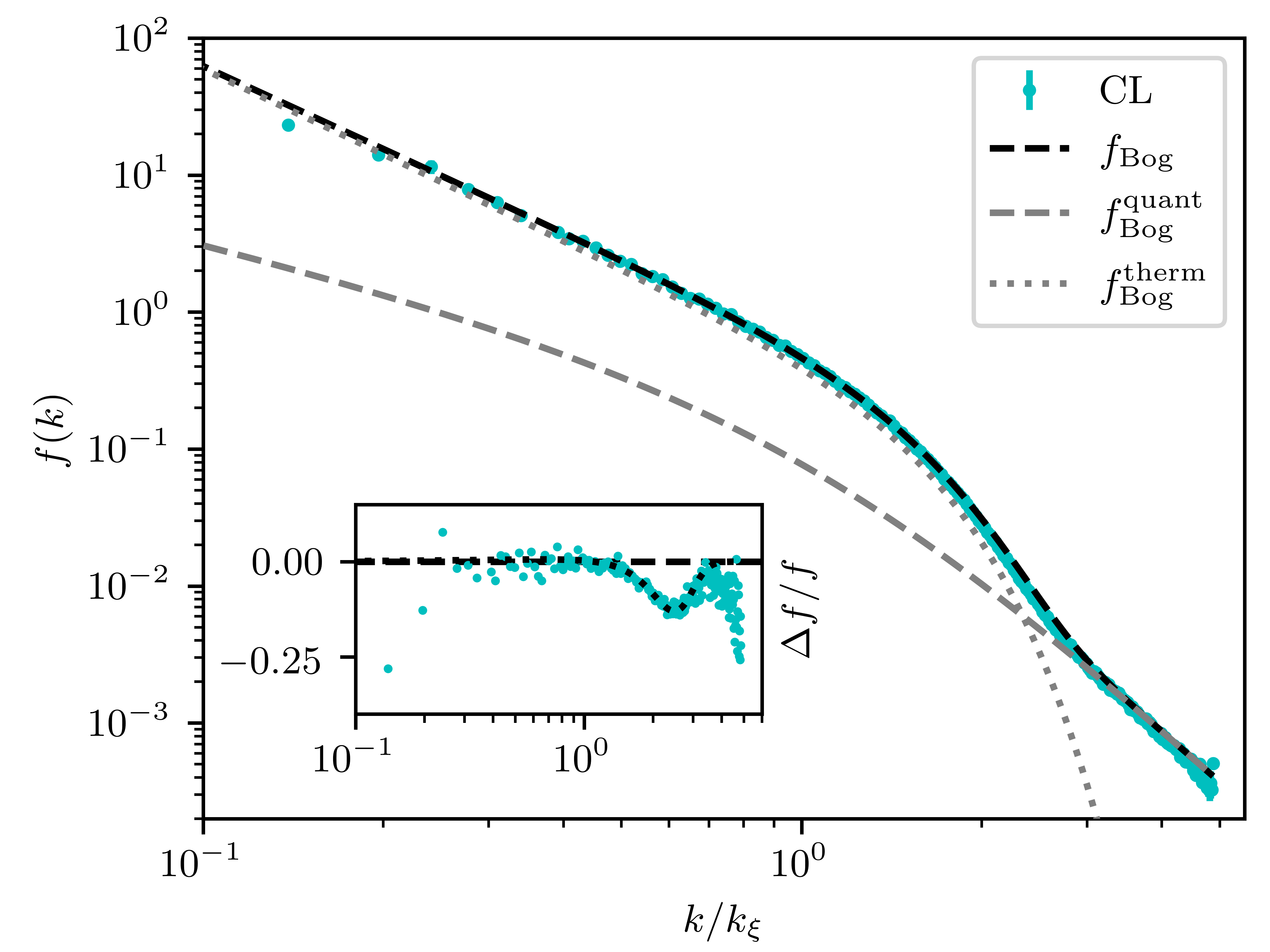}
	\caption{Momentum spectrum of the interacting single-component gas below the phase transition, at temperature  $T=0.625\,a_\mathrm{s}^{-1}$, chemical potential $\mu=0.5\,a_\mathrm{s}^{-1}$, and for a coupling $g=0.1\,a_\mathrm{s}^2$, corresponding to a diluteness $\eta\approx0.56\cdot10^{-3}$.
	It is obtained from CL dynamics on a $64^{3}\times (N_{\tau}=32)$ lattice, averaged over $4$ runs starting from a state with constant density $\rho_{0}=\mu/g=5\,a_\mathrm{s}^{-3}$. The CL process leads to a slightly larger condensate density  $\rho_{0}=f(0)/\mathcal{V}=(5.0309\pm0.0002)\,a_\mathrm{s}^{-3}$, corresponding to $f(0)\approx1.32\cdot10^{6}$, while the condensate depletion becomes $\rho'\equiv\rho-\rho_0=(0.0149\pm0.0001)\,a_\mathrm{s}^{-3}$.
	Momenta are given in units of the healing momentum $k_\xi=\sqrt{2mg\rho_0}=0.71\,a_\mathrm{s}^{-1}$.
	The dashed black line represents the spectrum \eq{bogol_spectrum} predicted by Bogoliubov theory, consisting of a thermal (dotted gray line) and a quantum depletion (dashed grey line) part.
	The inset shows the relative deviation  $(\Delta f/f)^\mathrm{CL}_\mathrm{Bog}=(f_\mathrm{CL}-f_\mathrm{Bog})/f_\mathrm{Bog}$ of the CL data from the Bogoliubov spectrum. 
	The dotted black line indicates an estimate of the deviation $(\Delta f/f)^{N_{\tau}}_{\mathrm{Bog}}=[f_\mathrm{Bog}({k};N_{\tau})-f_\mathrm{Bog}({k})]/f_\mathrm{Bog}({k})$ of the finite-$N_{\tau}$ from the $N_{\tau}\to\infty$ version of the Bogoliubov spectrum, cf.~the main text.
	 }
	\label{fig:belowtrans}
\end{figure}

When starting instead from a spatially uniform, non-vanishing field configuration given by the semi-classical mean-field value $\psi(\tau,\mathbf{x})\equiv\sqrt{\mu/g}$, the CL process leads to only a slight modification, which consists in a thermal depletion of the condensate mode and the buildup of the corresponding distribution of momentum excitations shown in \Fig{belowtrans}.
%==============================================================
%==============================================================

From Bogoliubov mean-field theory one expects the momentum distribution of the single-component gas to be \cite{Pitaevskii2016bose}
\begin{align}
	\label{eq:bogol_spectrum}
	 f_{\mathrm{Bog}}(\mathbf{k})
	 &=\frac{1+2v_{\mathbf{k}}^{2}}{\mathrm{e}^{\,\beta\omega_{\mathrm{Bog}}(\mathbf{k})}-1}
	 +v_{\mathbf{k}}^{2}
	 \,,
\end{align}
where
\begin{align}
	\label{eq:bogol_v}
	 v_{\mathbf{k}}
	 &=\sqrt{\frac{1}{2}\left[\frac{\varepsilon(\mathbf{k})+g\rho_0}{\omega_{\mathrm{Bog}}(\mathbf{k})}-1\right]}
	 \,,
\end{align}
in which $\rho_{0}$ denotes the condensate density
\begin{align}
	\label{eq:omegaBog}
	\rho_{0}
	&=\rho - \mathcal{V}^{-1}\sum_{\mathbf{k}\not=0}f_{\mathrm{Bog}}(\mathbf{k})
	\,,
\end{align}
with spatial volume $\mathcal{V}$. 
The Bogoliubov dispersion of the elementary excitations is given in terms of the free dispersion $\varepsilon(\mathbf{k})$, \Eq{freegasdisp}, by
\begin{align}
	\label{eq:omegaBog2}
	\omega_{\mathrm{Bog}}(\mathbf{k})
	&=\sqrt{\varepsilon(\mathbf{k})\left[\varepsilon(\mathbf{k})+2g\rho_0\right]}
	\,.
\end{align}
Note that the quasiparticle dispersion is sound like, $\omega_{\mathrm{Bog}}(\mathbf{k})\simeq c_\mathrm{s}|\mathbf{k}|$, at momenta below the healing-length scale $k_{\xi}=\sqrt{2mg\rho_{0}}$.

The second contribution to \eq{bogol_spectrum}, $v_{\mathrm{k}}^{2}$, gives rise to the quantum depletion, i.e., it accounts for all particles that are not in the condensate mode due to the interactions even at $T=0$.
The first term, which is proportional to the thermal occupancy of the quasiparticles, approximated as being non-interacting, accounts for the fraction of particles, which are non-condensed due to thermal excitations.

In \Fig{belowtrans}, we show the spectrum \eq{bogol_spectrum} as a black dashed line, which hardly deviates from our numerical data, as reflected by the relative deviation $(\Delta f/f)^\mathrm{CL}_\mathrm{Bog}=(f_\mathrm{CL}-f_\mathrm{Bog})/f_\mathrm{Bog}$ depicted in the inset.
The grey dotted and dashed lines indicate the contributions to the analytic spectrum \eq{bogol_spectrum}, which account for the thermal and quantum depletion of the condensate, respectively.

We estimate the deviation caused by a finite $N_\tau$ by replacing the thermal distribution of non-interacting quasiparticles in \Eq{bogol_spectrum} by the finite-$N_{\tau}$ expression, as we did this for the ideal gas,
\begin{align}
	\frac{1}{\mathrm{e}^{\,\beta\omega_{\mathrm{Bog}}(\mathbf{k})}-1}
	\to\sum_{n=0}^{N_{\tau}-1}\frac{1}{{N_\tau}(\mathrm{e}^{2\pi\mathrm{i} n/{N_\tau}}-1)+\beta\omega_{\mathrm{Bog}}(\mathbf{k})}
	\,.
\end{align}
As can be inferred from the inset of \Fig{belowtrans}, this captures the residual deviation between the CL and Bogoliubov distributions up to momenta above which the quantum depletion becomes more relevant than the thermal one.

%==============================================================
\begin{figure}	
	\includegraphics[width=0.95\columnwidth]{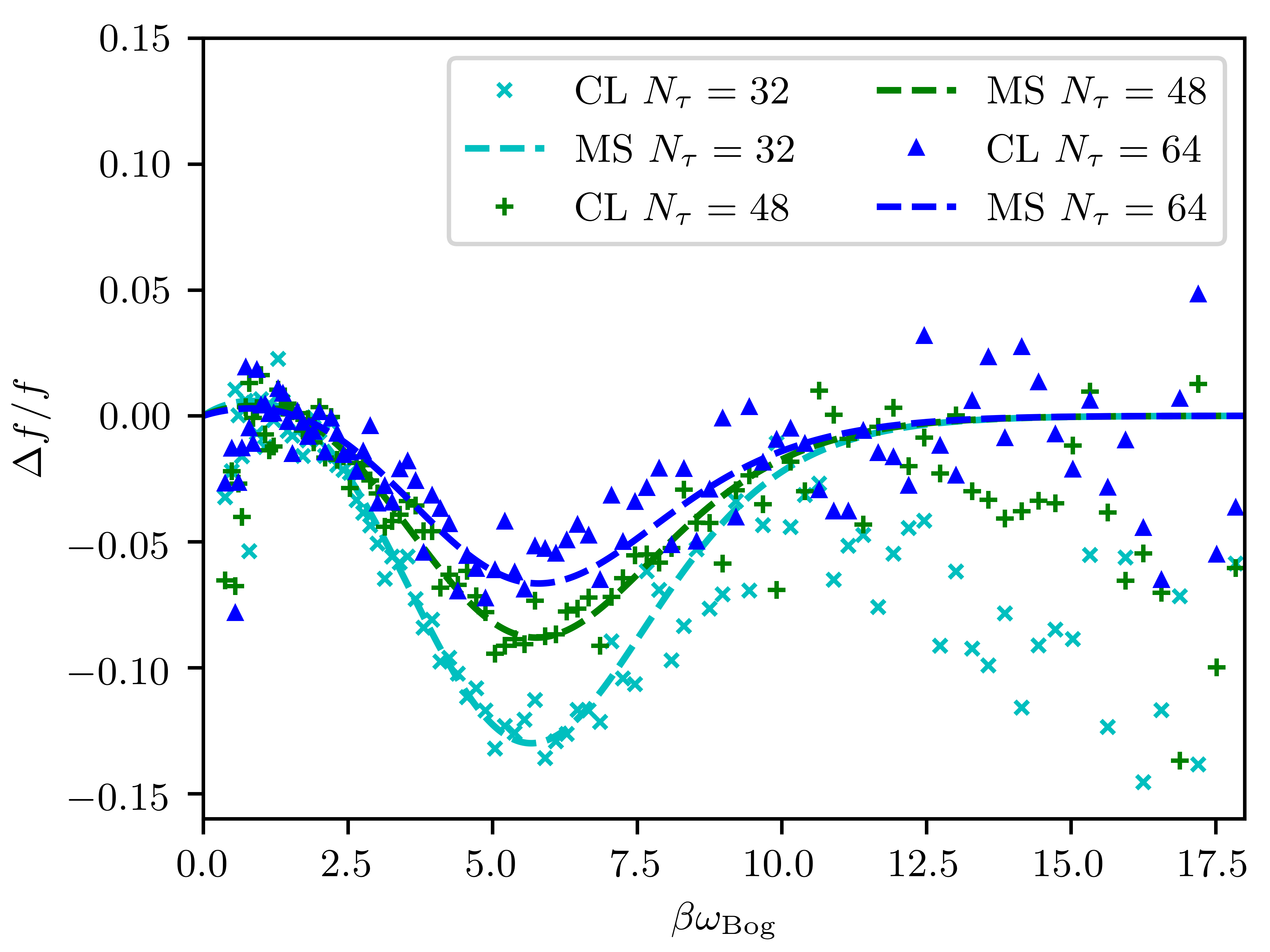}
	\caption{Comparison between the relative deviation $(\Delta f/f)^\mathrm{CL}_\mathrm{Bog}=(f_\mathrm{CL}-f_\mathrm{Bog})/f_\mathrm{Bog}$ of the numerical and Bogoliubov, \eq{bogol_spectrum}, distributions (CL, points) and the finite-size deviation $(\Delta f/f)^{N_{\tau}}_{\mathrm{Bog}}=[f_\mathrm{Bog}({k};N_{\tau})-f_\mathrm{Bog}({k})]/f_\mathrm{Bog}({k})$ due to the truncation of the Matsubara series (MS, dashed lines), as a function of $\beta\omega_{\mathrm{Bog}}$.
	The data is obtained for the same parameters as in \Fig{belowtrans}, but for three different $N_{\tau}\in\{32,48,64\}$. Note that for better visibility we doubled the width of the momentum bins in comparison to \Fig{belowtrans}.
	}
	\label{fig:comparisonBog}
\end{figure}

%==============================================================
In \Fig{comparisonBog} we compare the deviations of the distributions obtained with CL and the Bogoliubov distribution with the respective deviations of the finite-$N_{\tau}$ estimate from the full Bogoliubov distribution, for three different $N_{\tau}\in\{32,48,64\}$.
This demonstrates that the Matsubara series truncation also affects the region of high momenta where $v_{k}^{2}$ dominates the spectrum.

%==============================================================
\begin{figure}	
	\includegraphics[width=0.95\columnwidth]{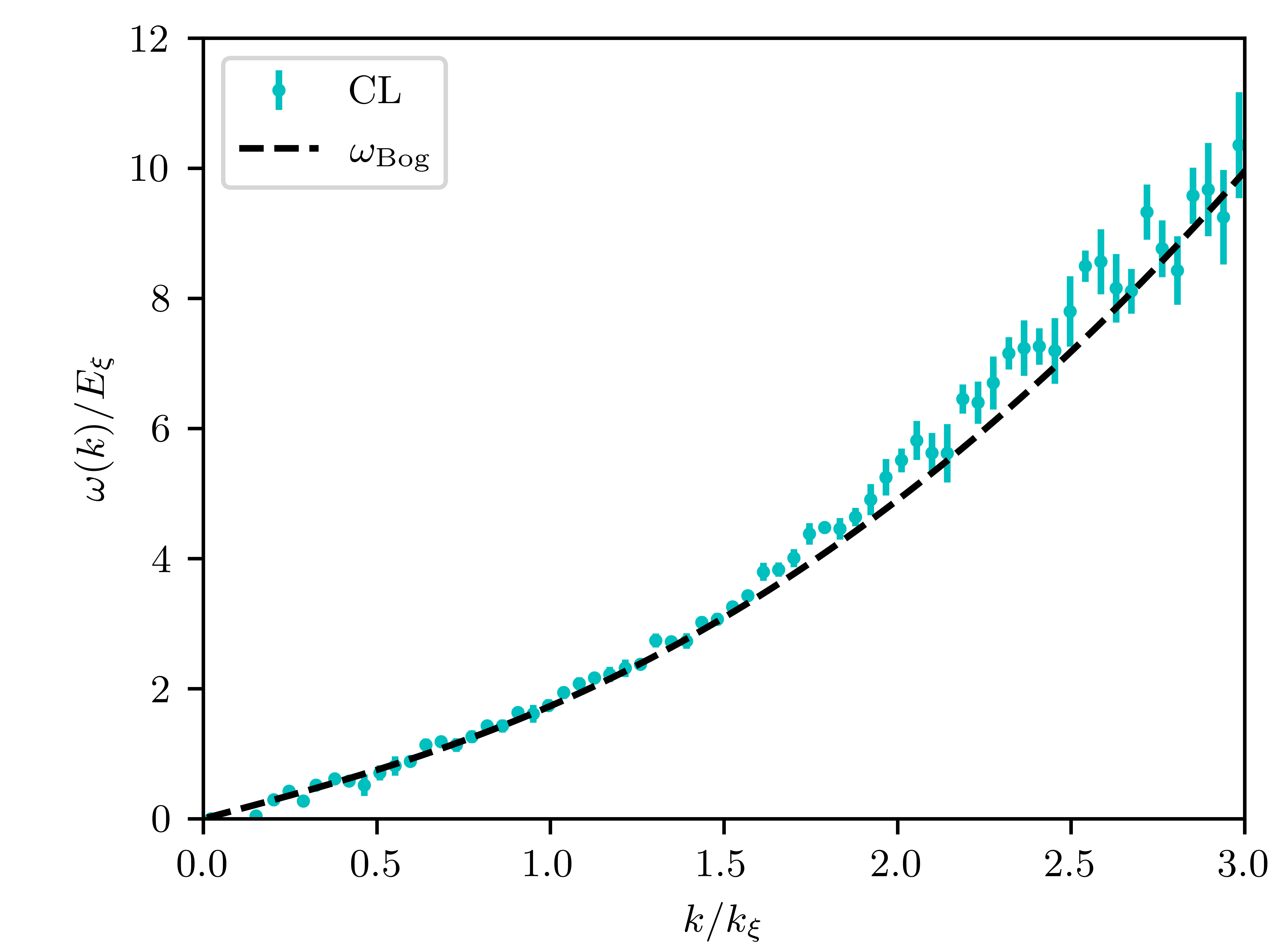}
	\caption{Dispersion of the interacting single-component gas below the phase transition in units of $E_\xi\equiv k_\xi^2/2m$, as obtained according to \Eq{disp} from the same data as used in \Fig{belowtrans} and averaged over the angular orientations of $\mathbf{k}$. Note that for better visibility we doubled the width of the momentum bins in comparison to \Fig{belowtrans}.
	For comparison we show the dispersion predicted by Bogoliubov theory as a black dashed line.
	}
	\label{fig:belowtrans_disp}
\end{figure}

\begin{figure}[th!]
	\includegraphics[width=0.95\columnwidth]{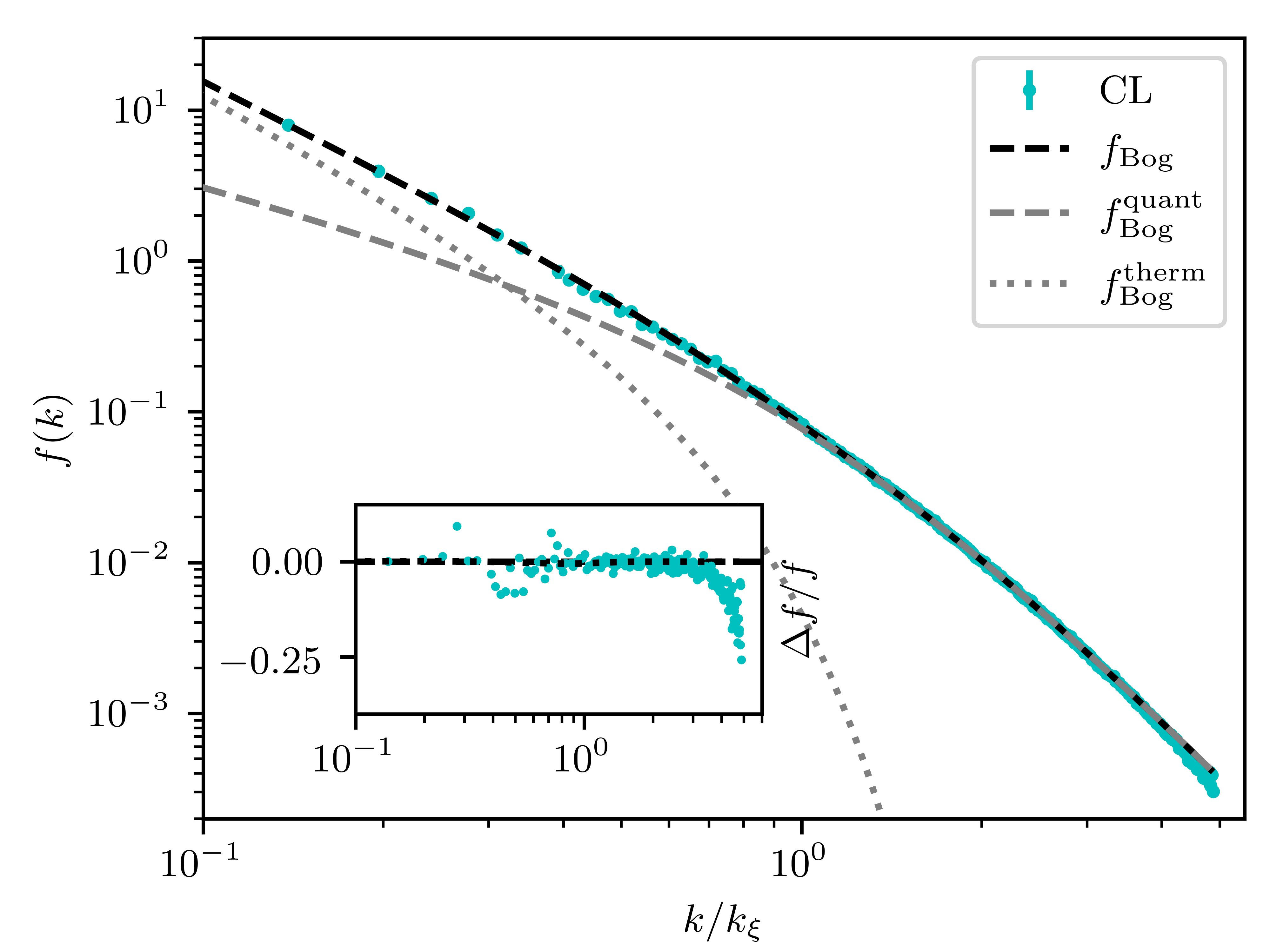}
	\caption{The same as in Fig. \ref{fig:belowtrans}, but for $T=0.15625\,a_\mathrm{s}^{-1}$ and $N_\tau=128$. At this low temperature, large parts of the spectrum are dominated by the quantum depletion, yet CL is still able to precisely describe the spectrum.
	}
	\label{fig:belowtrans_lowT}
\end{figure}
%==============================================================

Also the Bogoliubov dispersion, which we show in \Fig{belowtrans_disp}, is well reproduced by the CL simulations.

Finally, it is important to note that the CL method is not only able to describe the part of the momentum spectrum which is dominated by thermal excitations but also to reproduce the occupation numbers at high momenta where the spectrum is dominated by the quantum contribution. In order to further corroborate this observation, we also performed a simulation for a much smaller temperature, $T=0.15625\,a_\mathrm{s}^{-1}$, i.e., we increased $N_\tau$ to $128$. For such low temperatures, the spectrum is almost entirely dominated by the quantum depletion. Still, the spectrum is accurately reproduced by CL, see Fig. \ref{fig:belowtrans_lowT}. This gives hope that CL is a suitable tool for performing exact simulations also in the quantum regime.  
%======================================================================================
\subsubsection{Two-component system}
We have repeated the above simulations for the case of a two-component system, with U$(2)$ symmetric Hamiltonian \eq{Hamiltonian}.
From Bogoliubov theory (see \App{bogol} for a summary for the general U$(\mathcal{N})$ case), one obtains, as mentioned initially, besides the gapless elementary excitations of the total density with dispersion \eq{omegaBog}, an additional free Goldstone mode with dispersion given by \eq{freegasdisp} for each of the remaining $\mathcal{N}-1$ degrees of freedom.
For the case of general $\mathcal{N}$, the total momentum distribution reads
\begin{align}
	\label{eq:bogol_spectrumON}
	 f_{\mathrm{Bog},\mathrm{tot}}(\mathbf{k})
	 &= \sum_{a=1}^{\mathcal{N}}\langle|\psi_a(\mathbf{k})|^2\rangle
	 \nonumber\\
	 &=\frac{1+2v_{\mathbf{k},\mathcal{N}}^{2}}{\mathrm{e}^{\,\beta\omega_{\mathrm{Bog}}(\mathbf{k})}-1}
	 +v_{\mathbf{k},\mathcal{N}}^{2}
	 +\frac{\mathcal{N}-1}{\mathrm{e}^{\,\beta\varepsilon(\mathbf{k})}-1}
	 \,,
\end{align}
where
\begin{align}
	\label{eq:bogol_vON}
	 v_{\mathbf{k},\mathcal{N}}
	 &=\sqrt{\frac{1}{2}\left[\frac{\varepsilon(\mathbf{k})+\mathcal{N}g\rho_0}{\omega_{\mathrm{Bog}}(\mathbf{k})}-1\right]}
	 \,,
\end{align}

\begin{figure}
	\includegraphics[width=0.95\columnwidth]{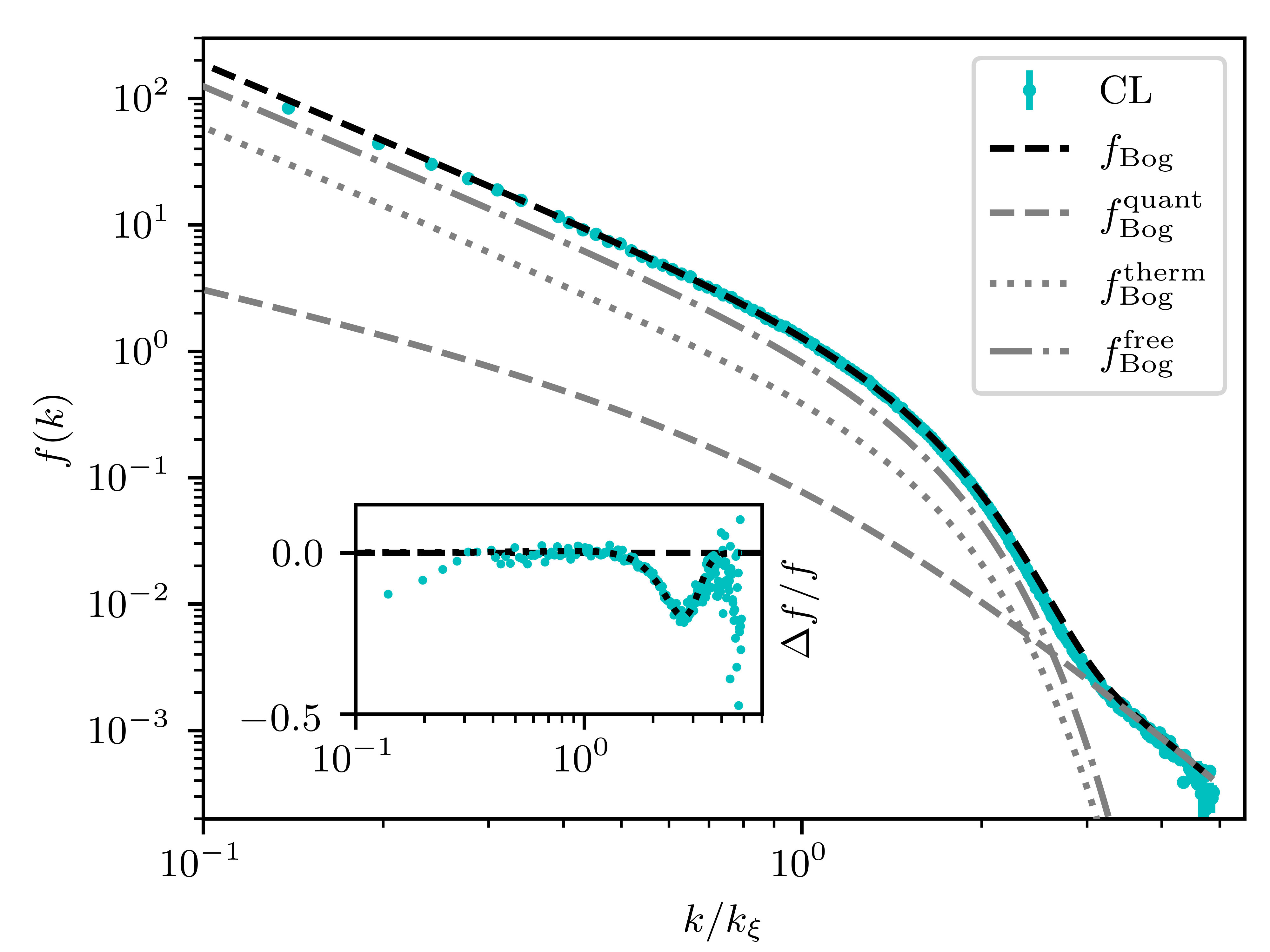}
	\caption{Momentum spectrum of the interacting two-component gas below the phase transition.
	The system is described by the U$(2)$-symmetric Hamiltonian \eq{Hamiltonian}, i.e., $\mathcal{N}=2$.
	 Parameters are, as before, $T=0.625\,a_\mathrm{s}^{-1}$, $\mu=0.5\,a_\mathrm{s}^{-1}$, and $g=0.1\,a_\mathrm{s}^2$, $N_{\tau}=32$, averaging over $3$ runs. 
	 The spectrum \eq{bogol_spectrumON} predicted by $\mathcal{N}$-component Bogoliubov theory is represented by the black dashed line. 
	 For comparison, we have also plotted the quantum (dashed gray line) and thermal (dotted gray line) contributions, as well as the contribution from the free Goldstone relative number excitations between the two internal components (dashed-dotted line). 
	 The inset shows the relative deviation $(\Delta f/f)^\mathrm{CL}_\mathrm{Bog,tot}=(f_\mathrm{CL}-f_\mathrm{Bog,tot})/f_\mathrm{Bog,tot}$ of the CL data from the Bogoliubov spectrum. 
	 The dotted black line shows an estimate of the deviation $(\Delta f/f)^{N_{\tau}}_{\mathrm{Bog,tot}}=[f_\mathrm{Bog,tot}({k};N_{\tau})-f_\mathrm{Bog,tot}({k})]/f_\mathrm{Bog,tot}({k})$ of the finite-$N_{\tau}$ from the $N_{\tau}\to\infty$ version of the Bogoliubov spectrum, cf.~the main text.
	}
	\label{fig:belowtransU2}
\end{figure}

We show the results of our CL simulations in \Fig{belowtransU2} and compare them to the total Bogoliubov distribution \eq{bogol_spectrumON}, as well as to the quantum and thermal contributions from the Bogoliubov and free modes, respectively.
%==============================================================
%==============================================================

%======================================================================================
\subsection{Interacting system at the transition}
\label{sec:IntBGatCriticality}
We finally would like to explore the CL approach to the dilute Bose gas close to the Bose-Einstein phase transition.
We start by considering the occupation number spectrum to show a pure Rayleigh-Jeans power law in the IR as a signature of the phase transition, before we move on to determining the shift of the critical temperature due to interactions in the real Bose gas.
%======================================================================================
\subsubsection{Rayleigh-Jeans scaling}
At the transition, we expect the occupation number to show Rayleigh-Jeans scaling,
\begin{align}
	f(\mathbf{k})\sim \frac{1}{|\mathbf{k}|^2}
	\,,
\end{align}
for small $|\mathbf{k}|$ in the infinite-volume limit. 
In principle, this could provide a method for determining the transition point, which, in practice, however, is suited only for obtaining a rough estimate of where the phase transition occurs. 
On the one hand, statistical errors are in general large for the relevant small momentum modes; 
on the other hand, in finite-size systems,  even at the transition, Rayleigh-Jeans scaling is not expected to be realized down to vanishing momenta.

As reported in the previous section, the formation of a condensate by the Langevin process can take unrealistically long computation times. 
Hence, for simulations below the transition, a condensate must be seeded by choosing a non-vanishing zero-mode population in the initial configuration for the Langevin process. 
However, since we do not know a priori at which chemical potential condensation occurs, it is most convenient to approach the critical point from the non-condensed phase. 
Nonetheless, the fact that the Langevin equilibration time strongly increases for a condensed state gives us a further (numerical) indication for determining the transition point.

We choose a temperature $T=1.25\,a_\mathrm{s}^{-1}$ and a coupling $g=0.5\,a_\mathrm{s}^2$, resulting in a density $\rho\approx0.08\,a_\mathrm{s}^{-3}$ and thus a diluteness of $\eta\approx 0.8\cdot 10^{-3}$. 
From HF theory, one obtains the transition to occur at $\mu=2g\rho_\mathrm{c}^0$, with $\rho_\mathrm{c}^0$ the critical density in the free system. 
This amounts to $\mu_\mathrm{c}^{\text{HF}}/T=0.066$ for chosen temperature and coupling.

%==============================================================
\begin{figure}[t]
	\includegraphics[width=0.95\columnwidth]{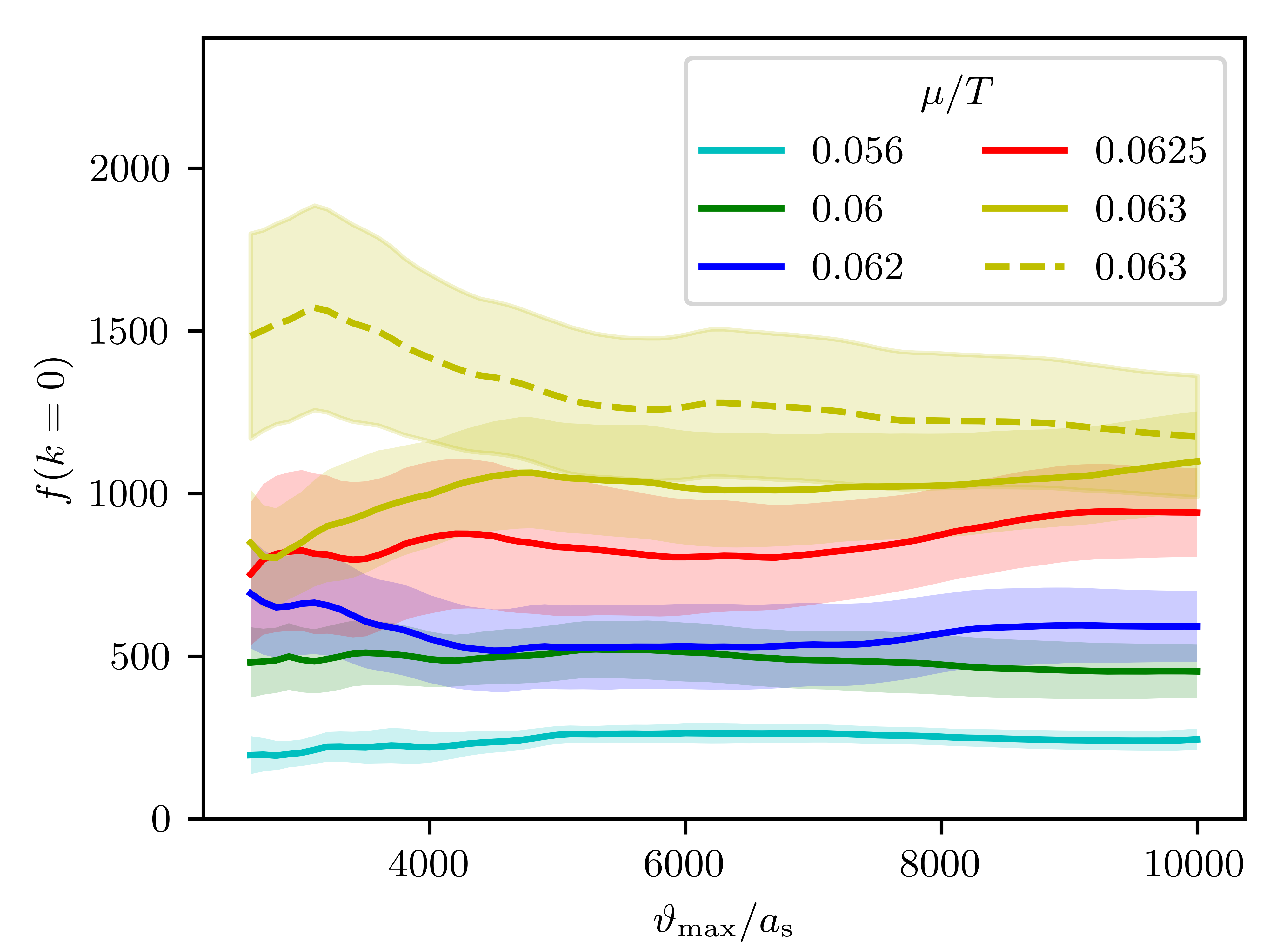}
	\caption{Langevin evolution of the zero mode occupancy $f(0)$ averaged over Langevin times $\vartheta=\vartheta_{0},\dots,\vartheta_{\mathrm{max}}$, cf.~\eq{f0CLaverage}, as a function of $\vartheta_{\mathrm{max}}$, with $\vartheta_0=2.5\cdot10^3\,a_\mathrm{s}$, for five different chemical potentials $\mu$ close to the transition, with chemical potential decreasing from the uppermost to the lowermost curve, at a temperature  $T=1.25\,a_\mathrm{s}^{-1}$, coupling $g=0.5\,a_\mathrm{s}^2$.
	The data is obtained on a $64^{3}\times (N_{\tau}=16)$ lattice and the resulting densities $\rho\approx0.08\,a_\mathrm{s}^{-3}$ correspond to a diluteness $\eta\approx0.8\cdot10^{-3}$.
	For $\mu/T=0.056$, $0.06$, $0.062$, $0.0625$, no seed of the zero mode was given. 
	For $\mu/T=0.063$, we simulated with a seed of $f(0)=0$ (solid line) and  $f(0)\approx1475$ ($\psi=0.075\,a_\mathrm{s}^{-3/2}$) (dashed line). 
	Error bands are obtained from the variance of $10$ independent runs.
	}
	\label{fig:timeevol}
\end{figure}
%==============================================================
We perform simulations for five values of the chemical potential between $\mu/T=0.056$ and $\mu/T=0.063$. 
The Langevin time evolution of the accumulated average of the zero-mode occupancy, 
\begin{align}
	f(0)
	\equiv\langle f(0,\vartheta)\rangle_{\vartheta_0}^{\vartheta_{\mathrm{max}}}
	=\frac{1}{\vartheta_\text{max}-\vartheta_0}\int\limits_{\vartheta_0}^{\vartheta_\text{max}}d\vartheta \,f(k=0,\vartheta)
	\,,
	\label{eq:f0CLaverage}
\end{align}
is shown in \Fig{timeevol}, as a function of the maximum Langevin time $\vartheta_{\mathrm{max}}$ and for $\vartheta_0=2.5\cdot10^3\,a_\mathrm{s}$.
While we observe a fast convergence in $\vartheta_{\mathrm{max}}$ for $\mu/T=0.056$ and $0.06$, i.e.~still rather far away from the transition, the convergence of $f(0)$ appears to be more and more slowed down for $\mu/T=0.062$, $0.0625$, and $0.063$, especially for the latter value of the chemical potential, giving a first hint that the system approaches the transition.
In order to make sure that our results are reliable and not biased by insufficient simulation time, we performed additional runs with a non-zero seed of $\psi=0.075\,a_\mathrm{s}^{-3/2}$, corresponding to $f(0)\approx1475$, for the case of the largest chemical potential, $\mu/T=0.063$. 
The error bands indicate the variance over $10$ independent runs and thus illustrate the demand for statistics in the zero mode.
The long-time zero-mode occupancy appears to converge for all chemical potentials and especially the two simulations with different seed for $\mu/T=0.063$ are consistent with each other within the error bands. 
All runs used in the remainder of this chapter have been performed without a seed in the zero-mode.

%==============================================================
\begin{figure}
	\includegraphics[width=0.95\columnwidth]{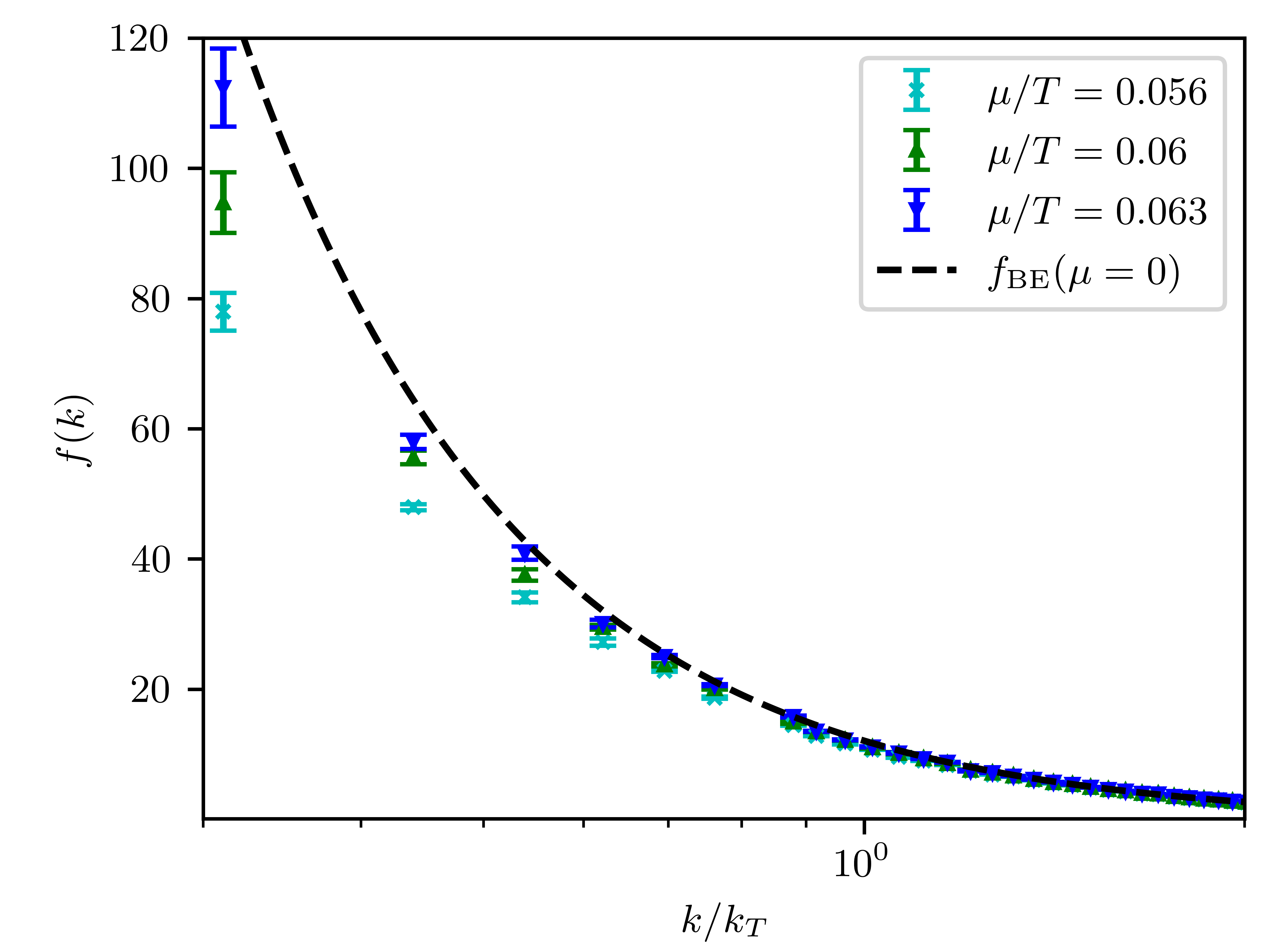}
	\caption{Infrared part of the momentum spectrum of the interacting single-component gas, for three different chemical potentials $\mu$ close to and below the transition, obtained by averaging up to $\vartheta_{\mathrm{max}}=10^{4}\,a_\mathrm{s}$. 
	All other parameters are as in \Fig{timeevol}. 
	Note that here only the $k$-axis carries a log scale for better visibility. 
	}
	\label{fig:acrosstrans}
\end{figure}
%==============================================================
The full spectra $f(k)\equiv\langle f(k,\vartheta)\rangle_{\vartheta_0}^{\vartheta_{\mathrm{max}}}$ for three of the five configurations shown in \Fig{timeevol} are depicted in \Fig{acrosstrans}, as obtained from averaging up to $\vartheta_{\mathrm{max}}=10^{4}\,a_\mathrm{s}$. 
They approach the Bose-Einstein distribution for $\mu=0$, while Rayleigh-Jeans scaling at low $k$ is not expected to be reached exactly in the finite-volume case.

%======================================================================================
\subsubsection{Shift of the critical temperature}
Let us now turn to the determination of the shift of the transition temperature due to interactions. 
At leading order in the dimensionless interaction strength, set by the diluteness $\eta$, \Eq{diluteness}, the shift scales as $\eta^{2/3}$  \cite{Baym1999transition}, and to next-to-leading order in $\eta$, it takes the form \cite{Arnold2001tc}
\begin{align}
	\label{eq:shiftc}
	\frac{\Delta T_\mathrm{c}}{T_\mathrm{c}^0}=c \,\eta^{2/3}
	+[c' \text{ln} ( \eta^{2/3} ) +c'' ]\,\eta^{4/3}\,,
\end{align}
where $c$, $c'$, and $c''$ are numerical constants, $T_\mathrm{c}^{0}=[\rho/\zeta(3/2)]^{2/3}2\pi/m$ is the critical temperature of the ideal Bose gas in $d=3$ dimensions and $\Delta T_\mathrm{c}\equiv T_\mathrm{c}-T_\mathrm{c}^0$ the shift of its value in the presence of interactions. 

\begin{figure}
	\includegraphics[width=0.95\columnwidth]{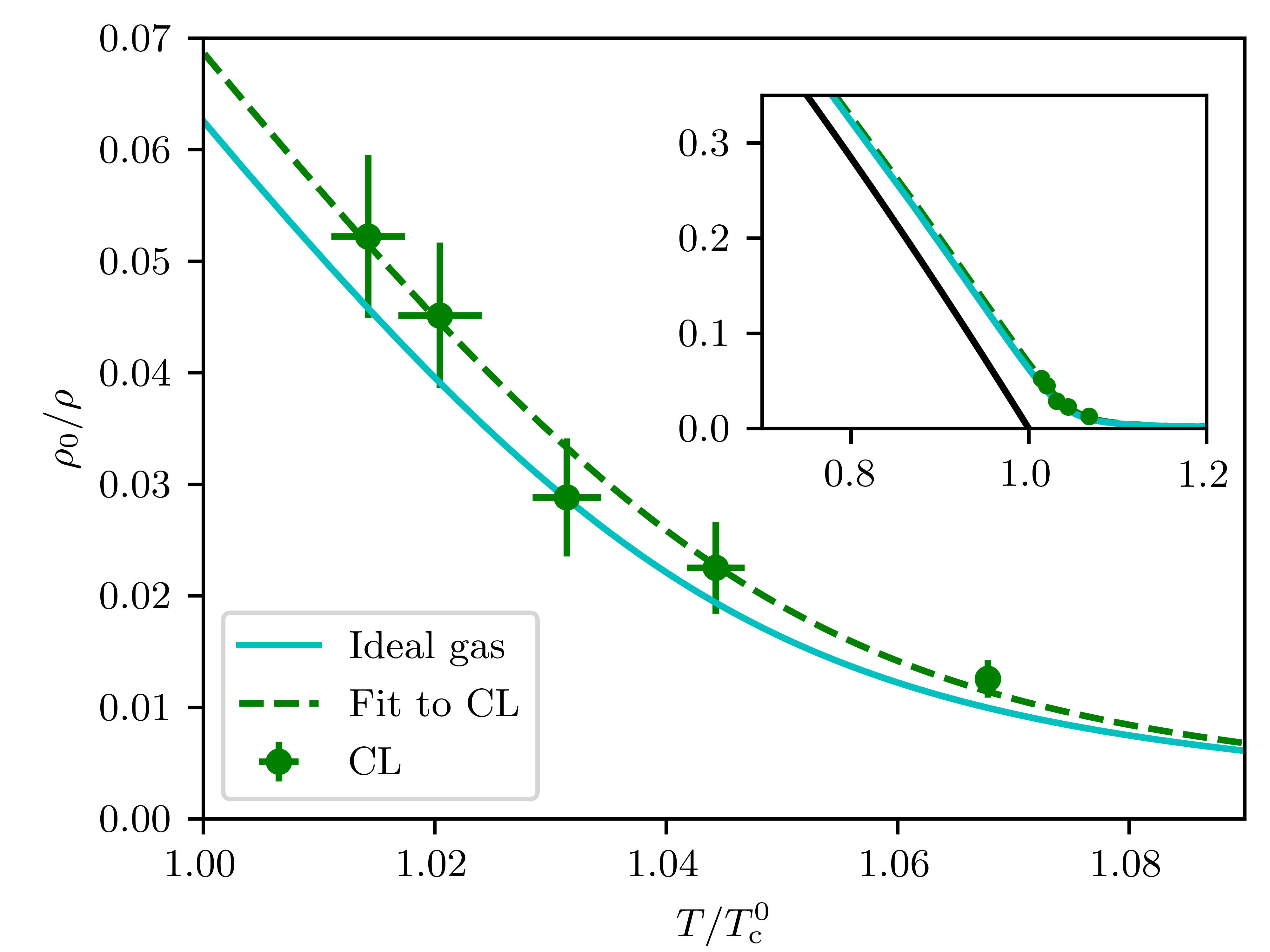}
	\caption{Condensate fraction $\rho_{0}/\rho$ as a function of $T/T_\mathrm{c}^{0}$, for a fixed temperature $T=1.25\,a_\mathrm{s}^{-1}$, coupling $g=0.5\,a_\mathrm{s}^2$ and system size $(64\,a_\mathrm{s})^{3}$. 
	The values obtained from the CL simulation for the five different chemical potentials $\mu$ from \Fig{timeevol} (green data points) are compared with the behavior of the ideal Bose gas (cyan solid line). 
	The dashed green line represents a fit of the ideal-gas behaviour, with shifted critical temperature,  to the simulation data as described in the main text, which gives $c=0.58\pm0.16$ quantifying the interaction-induced shift \eq{shiftc}.
	The inset depicts the same curves and data together with the near-critical condensate fraction in the thermodynamic limit (black solid line).
	}
	\label{fig:shift}
\end{figure}

While, in principle, a thorough determination of the transition point requires carefully extrapolating to the infinite-volume limit, for the case of a weakly interacting gas we can employ a somewhat simpler approach. 
Consider the condensate fraction $\rho_{0}/\rho$ in a free gas as a function of $T/T_\mathrm{c}^0$, for a fixed temperature $T$, and system size $\mathcal{V}=L^{3}$, and thus fixed ratio $\ell=L/\lambda_T$ of the system size $L$ and the thermal wave length $\lambda_{T}= \sqrt{2\pi/mT}$. 
Varying the chemical potential $\mu$ allows tuning the total density $\rho=N/\mathcal{V}$ as well as the zero-mode density $\rho_{0}=f(0)/\mathcal{V}$, and the critical temperature $T_\mathrm{c}^{0}$, such that
\begin{align}
	{T}/{T_\mathrm{c}^{0}}
	&=\ell^{2}\left(\frac{\zeta(3/2)}{N}\right)^{2/3}
	\,,\\
	{\rho_{0}}/{\rho}
	&=\frac{z}{1-z}N^{-1}
	\,,
\end{align}
where $z=\exp(\beta\mu)$ is the fugacity.
The total particle number as a function of $z$ in the continuum limit can be determined as the sum 
\begin{align}
	N
	=\sum_{m_{x}m_{y}m_{z}}
	\frac{1}{z^{-1}\exp\left[{\pi}{\ell^{-2}}(m_{x}^2+m_{y}^2+m_{z}^2)\right]-1}
	\,,
\end{align}
where the sums are performed over $m_{i}=-M_{i},\dots,M_{i}$, with $M_{x}=M_{y}=M_{z}$ chosen large enough to ensure convergence. 
The resulting dependence of the ideal-gas condensate fraction $\rho_{0}/\rho$ on $T/T_\mathrm{c}^{0}$ is shown in \Fig{shift} as a solid cyan line.
For comparison, the inset shows the same curve over a wider range near the critical temperature, together with the critical scaling $\rho_{0}/\rho=1-(T/T_\mathrm{c}^{0})^{3/2}$ (black solid line), which results in the thermodynamic limit $\mathcal{V}\to\infty$.

For the weakly interacting Bose gas, we make use of the approximation, that the functional form of the condensate fraction as a function of $T$ in units of the critical temperature is the same as in the non-interacting case, albeit with a shifted critical temperature, i.e.
\begin{align}
	\left[{\rho_{0}}/{\rho}\right]({T}/{T_\mathrm{c}^{0}})
	=\left[{\rho_{0}}/{\rho}\right]_\mathrm{free}(\alpha{T}/{T_\mathrm{c}^{0}})
	\,,
\end{align} 
with $\alpha=T_c^0/T_c$ and thus $\Delta T_c/T_c^0=1/\alpha-1$. 
Within errors, this is confirmed by our data.
For each of the chemical potentials chosen in our CL simulations, we determine the total density $\rho$ from the sum over the occupation number spectrum over all points of the momentum-time lattice, which we correct by the difference between the continuum and lattice densities determined for the ideal gas at the same temperature and chemical potential, see \App{ExtrDensity} for details.
In this way, we determine condensate fractions and corresponding values of ${T}/{T_\mathrm{c}^{0}}$ for the five different near-critical $\mu\in\{0.063,\dots,0.056\}$ considered above, which we can compare, cf.~\Fig{shift}, with the behaviour of the ideal gas in the continuum.

A least-squares fit of the ideal-gas curve to the simulation data (dashed green line) yields an interaction-induced shift of the critical temperature of
\begin{align}
	\label{eq:Tshiftnum}
	\Delta T_c/T_c^0
	=1/\alpha-1
	=0.00497\pm 0.00138\,.
\end{align}
Possible finite-size corrections to this result are discussed in App. \ref{app:finitesize}. For the density at $\mu/T=0.063$ the diluteness becomes $\eta=7.95\cdot10^{-4}$.
As a consequence, neglecting the $\mathcal{O}(\eta^{4/3})$-corrections in \eq{shiftc}, the constant $c$ in \Eq{shiftc} takes the value
\begin{align}
	\label{eq:cshift}
	c=0.58\pm 0.16\,.
\end{align}
%
%==============================================================
%==============================================================
%
The determination of the constant $c$ has a long and controversial history \cite{Stoof1992nucleation,Gruter1997a.PhysRevLett.79.3549,Holzmann1999bose,Reppy2000density,Arnold2000t,Arnold2001bec,Kastening2004bose,Nho2004bose}, see \cite{Andersen2004a} for a review. 
Within the errors, our result \eq{cshift} is in good agreement with the value $c=0.7$ (no error provided) from \cite{Holzmann1999bose}, while it is larger than $c=0.34\pm0.03$ from the PIMC simulations of \cite{Gruter1997a.PhysRevLett.79.3549}  and  smaller than the average of more recent results which accumulate near $c\approx 1.3$, compare $c=1.32\pm0.02$ from \cite{Arnold2001bec}, $c=1.29 \pm 0.05$ from \cite{Kashurnikov2001critical} (Monte Carlo simulation of classical $O(2)$ field theory), $c=1.27\pm0.11$ from \cite{Kastening2004bose} (variational perturbation theory),  and $c=1.32\pm0.14$ from \cite{Nho2004bose} (PIMC simulation). 
Experiments with helium in Vycor glasses have found $c\approx5.1$ \cite{Reppy2000density}, while this has been disputed in \cite{Arnold2000t}. 

We also note that in \cite{Arnold2001tc} the constants $c'$ and $c''$, which quantify the corrections to \eqref{eq:shiftc} of order $\eta^{4/3}$ were determined. 
The authors found $c=1.32\pm 0.02$, $c'=19.7518$ and $c''=75.7\pm0.4$, which for our diluteness $\eta=7.95\cdot10^{-4}$ yield an $\mathcal{O}(\eta^{4/3})$-correction to the shift $[c' \text{ln} ( \eta^{2/3} ) +c'' ]\,\eta^{4/3}=-0.00135\pm 0.00003$ and a total shift of $\Delta T_\mathrm{c}/T_\mathrm{c}^0=0.00998\pm 0.00017$, which is to be compared with \eq{Tshiftnum}.
In order to distinguish numerically the leading-order and next-to-leading order contributions would require an analysis for a range of different densities or interaction strengths and thus dilutenesses, which is beyond the scope of the present work.

In summary, our analysis demonstrates that the CL method gives access to a beyond mean-field quantity (in the mean-field approximation, there is no shift of the transition temperature), consistent with previous results by order of magnitude.

%======================================================================================
%======================================================================================
\section{Conclusion and Outlook}
We have shown that it is possible to evaluate the coherent-state path integral of the interacting Bose gas from first principles by means of the complex Langevin (CL) method in experimentally relevant coupling regimes. 
We found that the spectra and dispersions obtained via the Hartree-Fock and Bogoliubov approximations are well reproduced above and below the transition, respectively. 
By determining the shift of the critical temperature $\Delta T_\mathrm{c}$ due to interactions, we could show that the method is in principle capable of providing corrections beyond mean-field approximations. 
This opens a perspective on possibilities of using CL for ab initio simulations of interacting Bose gases in a wide range of experimental settings.

A remaining potential challenge is using the method in the regime of strong coupling strengths. 
Here we find CL to perform very well in the experimentally relevant regime of dilute gases, with diluteness parameter $\eta\sim10^{-3}$, without the need for additional improvements such as adaptive time step size or regulators. 
Notwithstanding this, choosing a diluteness of one order of magnitude higher ($\eta\sim 10^{-2}$, corresponding to a by two orders higher particle density), we encounter runaways into the complex plane of field values, which are typical for the method in parameter regimes, in which the so-called sign problem prevails. 
While such highly dense Bose gases are experimentally difficult to realize, it remains an interesting task beyond the scope of our work to determine the limits up to which the CL method leads to meaningful results in an efficient manner.

For dilute systems, the method promises to be useful in many different contexts, e.g., trapped systems beyond the local density approximation \cite{Goldman1981atomic}, two-dimensional gases near the BKT transition \cite{Foster2010a,Prokofiev2001critical}, or spinor gases beyond the U$(\mathcal{N})$-symmetric case \cite{Stamper-Kurn2013a.RevModPhys.85.1191}, for exploring the phase diagram in comparison with mean-field predictions \cite{Kawaguchi2012finite,Kawaguchi2012a.PhyRep.520.253}, or the exploration of anomalous averages and their interplay with the gaplessness of the spectrum \cite{Yukalov2006a.PhysRevA.73.063612}.
In principle, one may consider CL also for simulating real-time path integrals and thus time evolving systems.
This, however, poses a far more challenging sign problem, so far providing access to short-time evolution only  \cite{Berges2007lattice}.

%======================================================================================
%======================================================================================
\section{Acknowledgments}
The authors thank Felipe Attanasio,  Marc Bauer, Iacopo Carusotto, Stefanie Czischek, Pjotr Deuar, Martin G\"arttner, Christof Gattringer, Lukas Kades, Jan Pawlowski and Axel Pelster for discussions and collaboration on related topics. 
This work is supported by the Deutsche Forschungsgemeinschaft (DFG, German Research Foundation) under Germany's Excellence Strategy EXC 2181/1- 390900948 (the Heidelberg STRUCTURES Excellence Cluster), under SFB 1225 ISOQUANT - 273811115, as well as grant GA677/10-1.
The authors furthermore acknowledge support by the state of Baden-W\"urttemberg through bwHPC and the German Research Foundation (DFG) through grant INST 35/1134-1 FUGG (MLS-WISO cluster) and grant no INST 40/575-1 FUGG (JUSTUS 2 cluster).
\bigskip

%======================================================================================
%======================================================================================
%\clearpage
\appendix
\begin{center}
	\textbf{APPENDIX}
\end{center}
%
%======================================================================================
%======================================================================================
\section{Discrete Langevin equations}
\label{app:langeq}
In this appendix, we provide details of the discretization of the action \eq{action} and the complex Langevin equations  \eq{cl_eq} derived from it.

We express the complex Bose fields in terms of its real and imaginary parts as 
\begin{align}
	\psi_a\equiv\varphi_a+\mathrm{i}\chi_a
	\,. 
\end{align}
Discretizing these components on the $N_\mathrm{s}^{3}\times N_{\tau}$ lattice, $\varphi_{a}(\tau,\mathbf{x})=\varphi_{a}(ia_{\tau},\mathbf{j}a_\mathrm{s})\equiv\varphi_{a,i,\mathbf{j}}$, etc. for $\chi_{a}$, the action \eq{action} can be written as 
\begin{align}
	\label{eq:latticeaction}
	S^\text{lat}
	= a_\mathrm{s}^3a_\tau &
	\sum_{i,\mathbf{j}}\Bigg\{\sum_a\bigg[
	\frac{\varphi_{a,i,\mathbf{j}}-\varphi_{a,i-1,\mathbf{j}}}{a_\tau}\varphi_{a,i,\mathbf{j}}
	+\frac{\chi_{a,i,\mathbf{j}}-\chi_{a,i-1,\mathbf{j}}}{a_\tau}\chi_{a,i,\mathbf{j}}
	\nonumber\\
	&\qquad\qquad
	+\mathrm{i}\frac{\varphi_{a,i-1,\mathbf{j}}\,\chi_{a,i,\mathbf{j}}-\varphi_{a,i,\mathbf{j}}\,\chi_{a,i-1,\mathbf{j}}}{a_\tau}
	\nonumber\\
	&-\frac{1}{2ma_\mathrm{s}^2}\bigg(
	\varphi_{a,i+1,\mathbf{j}}\,\Delta^\text{lat}\varphi_{a,i,\mathbf{j}}
	+\chi_{a,i+1,\mathbf{j}}\,\Delta^\text{lat}\chi_{a,i,\mathbf{j}}
	\nonumber\\
	&\qquad\qquad+\mathrm{i}\varphi_{a,i+1,\mathbf{j}}\,\Delta^\text{lat}\chi_{a,i,\mathbf{j}}
	-\mathrm{i}\chi_{a,i+1,\mathbf{j}}\,\Delta^\text{lat}\varphi_{a,i,\mathbf{j}}\bigg)
	\nonumber\\
	&-\mu\,\bigg(\varphi_{a,i+1,\mathbf{j}}\,\varphi_{a,i,\mathbf{j}}
	+\chi_{a,i+1,\mathbf{j}}\,\chi_{a,i,\mathbf{j}}
	\nonumber\\
	&\qquad+\mathrm{i}\varphi_{a,i+1,\mathbf{j}}\,\chi_{a,i,\mathbf{j}}
	-\mathrm{i}\varphi_{a,i,\mathbf{j}}\,\chi_{a,i+1,\mathbf{j}}\bigg)\bigg]
	\nonumber\\
	&+\frac{g}{2}\,\bigg(\sum_a\big\{\varphi_{a,i+1,\mathbf{j}}\,\varphi_{a,i,\mathbf{j}}
	+\chi_{a,i+1,\mathbf{j}}\,\chi_{a,i,\mathbf{j}}
	\nonumber\\
	&\qquad+\mathrm{i}\varphi_{a,i+1,\mathbf{j}}\,\chi_{a,i,\mathbf{j}}
	-\mathrm{i}\varphi_{a,i,\mathbf{j}}\,\chi_{a,i+1,\mathbf{j}}\big\}\bigg)^2
	\Bigg\}
	\,,
\end{align}
where the index $i$ enumerates the imaginary time lattice sites with spacing $a_\tau$, the three-dimensional index vector $\mathbf{j}$  the spatial lattice sites, with discretization $a_\mathrm{s}$. 
$\Delta^\text{lat}$ is the Laplacian on the lattice, i.e.
\begin{align}
	\Delta^\text{lat}A_{a,i,\mathbf{j}}
	\equiv&\ 
	A_{a,i,\mathbf{j}+\mathbf{e}_x}+A_{a,i,\mathbf{j}-\mathbf{e}_x}+
	A_{a,i,\mathbf{j}+\mathbf{e}_y}+A_{a,i,\mathbf{j}-\mathbf{e}_y}
	\nonumber\\
	&\ +
	A_{a,i,\mathbf{j}+\mathbf{e}_z}+A_{a,i,\mathbf{j}-\mathbf{e}_z}-6A_{a,i,\mathbf{j}}
	\,,
	\label{eq:LaplaceLatt}
\end{align}
with $\mathbf{e}_{x,y,z}$ the unit vectors in three dimensions. 
Note that $\psi^*_a$ must be evaluated infinitesimally later than $\psi_a$, i.e.~at lattice point $i+1$ rather than at $i$, as can be seen from the construction of the coherent-state path integral where in each time step $\psi^*_a$ is evaluated with respect to the coherent state on the left while $\psi_a$ acts to the right, on the state one step earlier. 
This is important even in the analytical treatment where it ensures the right convergence in the complex Matsubara plane \cite{Altland2010a}.

The corresponding Langevin equations are obtained from \eq{latticeaction} by taking derivatives with respect to $\varphi$ and $\chi$, 
\begin{align}
	\label{eq:deriv1}
	\frac{\partial\varphi_{a,i,\mathbf{j}}}{\partial \vartheta}
	=&\ a_\mathrm{s}^3\,\left(
	\varphi_{a,i+1,\mathbf{j}}
	+\varphi_{a,i-1,\mathbf{j}}
	-2\varphi_{a,i,\mathbf{j}}
	+\mathrm{i}\chi_{a,i-1,\mathbf{j}}
	-\mathrm{i}\chi_{a,i+1,\mathbf{j}}
	\right)
	\nonumber\\
	&+\frac{a_\mathrm{s}a_\tau}{2m}\,\Delta^\text{lat}\bigg(
	\varphi_{a,i-1,\mathbf{j}}
	+\varphi_{a,i+1,\mathbf{j}}
	+\mathrm{i}\chi_{a,i-1,\mathbf{j}}
	-\mathrm{i}\chi_{a,i+1,\mathbf{j}}
	\bigg)
	\nonumber\\
	&+\mu a_\mathrm{s}^3a_\tau\,\bigg(
	\varphi_{a,i-1,\mathbf{j}}
	+\varphi_{a,i+1,\mathbf{j}}
	+\mathrm{i}\chi_{a,i-1,\mathbf{j}}
	-\mathrm{i}\chi_{a,i+1,\mathbf{j}}
	\bigg)
	\nonumber\\
	&-ga_\mathrm{s}^3a_\tau\,\bigg(
	\Psi_{i-1,\mathbf{j}}\,\varphi_{a,i-1,\mathbf{j}}
	+\Psi_{i,\mathbf{j}}\,\varphi_{a,i+1,\mathbf{j}}
	\nonumber\\
	&\qquad 
	+\mathrm{i}\Psi_{i-1,\mathbf{j}}\,\chi_{a,i-1,\mathbf{j}}
	-\mathrm{i}\Psi_{i,j,k,l}\,\chi_{a,i+1,\mathbf{j}}
	\bigg)+\eta(\vartheta)
\end{align}
and 
\begin{align}
	\label{eq:deriv2}
	\nonumber
	\frac{\partial\chi_{a,i,\mathbf{j}}}{\partial \vartheta}
	=&\ a_\mathrm{s}^3\,\left(
	\chi_{a,i+1,\mathbf{j}}
	+\chi_{a,i-1,\mathbf{j}}
	-2\chi_{a,i,\mathbf{j}}
	-\mathrm{i}\varphi_{a,i-1,\mathbf{j}}
	+\mathrm{i}\varphi_{a,i+1,\mathbf{j}}
	\right)
	\nonumber\\
	&+\frac{a_\mathrm{s}a_\tau}{2m}\,\Delta^\text{lat}\bigg(
	\chi_{a,i-1,\mathbf{j}}
	+\chi_{a,i+1,\mathbf{j}}
	-\mathrm{i}\varphi_{a,i-1,\mathbf{j}}
	+\mathrm{i}\varphi_{a,i+1,\mathbf{j}}\bigg)
	\nonumber\\
	&+\mu a_\mathrm{s}^3a_\tau\,\bigg(
	\chi_{a,i-1,\mathbf{j}}
	+\chi_{a,i+1,\mathbf{j}}
	-\mathrm{i}\varphi_{a,i-1,\mathbf{j}}
	+\mathrm{i}\varphi_{a,i+1,\mathbf{j}}\bigg)
	\nonumber\\
	&-ga_\mathrm{s}^3a_\tau\,\bigg(
	\Psi_{i-1,\mathbf{j}}\,\chi_{a,i-1,\mathbf{j}}
	+\Psi_{i,\mathbf{j}}\,\chi_{a,i+1,\mathbf{j}}
	\nonumber\\
	&\qquad 
	-\mathrm{i}\Psi_{i-1,\mathbf{j}}\,\varphi_{a,i-1,\mathbf{j}}
	+\mathrm{i}\Psi_{i,\mathbf{j}}\,\varphi_{a,i+1,\mathbf{j}}
	\bigg)+\eta(\vartheta)\,,
\end{align}
where
\begin{align}
	\Psi_{i,\mathbf{j}}
	\equiv&\sum_a\Big(
	\varphi_{a,i+1,\mathbf{j}}\,\varphi_{a,i,\mathbf{j}}
	+\chi_{a,i+1,\mathbf{j}}\,\chi_{a,i,\mathbf{j}}
	\nonumber\\
	&\qquad
	+\mathrm{i}\varphi_{a,i+1,\mathbf{j}}\,\chi_{a,i,\mathbf{j}}
	-\mathrm{i}\varphi_{a,i,\mathbf{j}}\,\chi_{a,i+1,\mathbf{j}}
	\Big)
	\,.
\end{align}
Note that both $\varphi$ and $\chi$ must now be taken to be \textit{complex} variables, such that the complex Langevin method requires to evolve \textit{four} real fields per component instead of two as in classical GPE simulations.
%
%======================================================================================
%======================================================================================
\section{\label{app:details}Numerical extraction of the observables}

%======================================================================================
\subsection{\label{app:spectrumanddispersion}Spectrum and dispersion on the lattice}
The translation of the occupation number $\langle a_\mathbf{k}^\dagger a_\mathbf{k}\rangle$ to the lattice gives $\langle\psi_{\mathbf{k},i+1}^*\psi_{\mathbf{k},i}\rangle$, where, as for the action, $\psi^*$ must be evaluated at the temporal lattice point $i+1$ and $\psi$ at $i$, following from the construction of the coherent state path integral, cf.~\App{langeq}.
When computing the dispersion, more care is in order.
In the operator formalism, we can write it as 
\begin{align}
	\omega(\mathbf{k})
	=\sqrt{-\frac{\partial_\tau\partial_{\tau'}\langle a^\dagger_{\mathbf{k}}(\tau)a_{\mathbf{k}}(\tau')\rangle|_{\tau=\tau'}}
		          {\langle a^\dagger_{\mathbf{k}}(\tau)a_{\mathbf{k}}(\tau')\rangle|_{\tau=\tau'}}}
	\,.
\end{align}
The discretization of $\partial_\tau\partial_{\tau'}\langle a^\dagger_{\mathbf{k}}(\tau)a_{\mathbf{k}}(\tau')\rangle|_{\tau=\tau'}$ on the temporal lattice is somewhat subtle. 
Since $\langle a^\dagger_{\mathbf{k}}(\tau)a_{\mathbf{k}}(\tau')\rangle|_{\tau=\tau'}$ translates to $\langle \psi_{\mathbf{k},i+1}^*\psi_{\mathbf{k},i}\rangle$, one is tempted to naively write $\langle (\psi_{\mathbf{k},i+2}^*-\psi_{\mathbf{k},i+1}^*)(\psi_{\mathbf{k},i+1}-\psi_{\mathbf{k},i})\rangle/a_\tau^2$. 
This, however, does \textit{not} correspond to the operator finite-time differences
\begin{align}
 \langle [a^\dagger_{\mathbf{k}}(\tau+a_\tau)-a^\dagger_{\mathbf{k}}(\tau)]
	[a_{\mathbf{k}}(\tau+a_\tau)-a_{\mathbf{k}}(\tau)]\rangle/a_\tau^2
	\,,
\end{align}
which contain an anti-time-ordered term, and thus cannot be simply translated to $\langle (\psi_{\mathbf{k},i+2}^*-\psi_{\mathbf{k},i+1}^*)(\psi_{\mathbf{k},i+1}-\psi_{\mathbf{k},i})\rangle/a_\tau^2$.

Hence, a better choice is to discretize the product of derivatives as
\begin{align}
	&\partial_\tau\partial_{\tau'}\langle a^\dagger_{\mathbf{k}}(\tau)a_{\mathbf{k}}(\tau')\rangle|_{\tau=\tau'}
	\nonumber\\
	&\approx \langle [a^\dagger_{\mathbf{k}}(\tau+a_\tau)-a^\dagger_{\mathbf{k}}(\tau)]
	[a_{\mathbf{k}}(\tau)-a_{\mathbf{k}}(\tau-a_\tau)]\rangle/a_\tau^2
	\,,
\end{align}
which contains time-ordered terms only and implies the discretization 
\begin{align}
	\label{eq:omegataudiscretized}
	\omega(\mathbf{k})
	=\sqrt{-\frac{\langle [\psi_{\mathbf{k},i+2}^*-\psi_{\mathbf{k},i+1}^*][\psi_{\mathbf{k},i}-\psi_{\mathbf{k},i-1}]\rangle}
		          {a_\tau^2\,\langle\psi_{\mathbf{k},i+1}^*\psi_{\mathbf{k},i}\rangle}}
\end{align}
of the dispersion, which we finally sum over $i$, in which is it homogeneous, to gain statistics.

%======================================================================================
\subsection{Momentum and integrals over momentum}

On a lattice of size $L=N_xa_\mathrm{s}=N_ya_\mathrm{s}=N_za_\mathrm{s}\equiv N_\mathrm{s}a_\mathrm{s}$, the kinetic energy associated with the momentum mode with indices $(j_x,j_y,j_z)$ is not given by $E_\text{kin}=(2\pi/L)^2(j_x^2+j_y^2+j_z^2)/2m$ as in the continuum. 
The discretization \eq{LaplaceLatt} of the Laplacian rather implies the energy to be sine-spaced, i.e.  
\begin{align}
	E_\text{kin}
	=\frac{2}{ma_\mathrm{s}^2}\left[
	\sin^2\left(\frac{\pi  j_x}{N_{x}}\right)
	+\sin^2\left(\frac{\pi j_y}{N_{y}}\right)
	+\sin^2\left(\frac{\pi j_z}{N_{z}}\right)\right]\,.
\end{align}
Accordingly, the modulus $k$ of the momentum associated with mode $(j_x,j_y,j_z)$ reads
\begin{align}
	k
	=\frac{2}{a_\mathrm{s}}\sqrt{
	\sin^2\left(\frac{\pi j_x}{N_{x}}\right)+
	\sin^2\left(\frac{\pi j_y}{N_{y}}\right)+
	\sin^2\left(\frac{\pi j_z}{N_{z}}\right)}
	\,.
\end{align}
In computing $f(k)$ and $\omega(k)$ we perform angular averages and accumulate the momenta into bins of width $(64\,a_\mathrm{s})^{-1}$. 
For the lowest 8 bins, which contain only a single value of $k$, we use this value as the $k$-coordinate in plots of functions of momentum, while for all other bins we choose the center of the bin.

Integrals over momentum space such as necessary for determining the mean density $\rho$ are evaluated as sums as 
\begin{align}
	\label{eq:rho}
	\rho
	=\frac{1}{L^3}\sum_{j_x j_y j_z}
	\left|
	\cos\left(\frac{\pi j_x}{N_{x}}\right)
	\cos\left(\frac{\pi j_y}{N_{y}}\right)
	\cos\left(\frac{\pi j_z}{N_{z}}\right)\right|
	f_{j_x j_y j_z}
	\,,
\end{align}
where we choose again $N_{x,y,z}\equiv N_\mathrm{s}=L/a_\mathrm{s}$ such that the sums run over $j_{x,y,z}=0,\dots N_\mathrm{s}-1$,  and where $f_{j_x j_y j_z}=\langle\psi_{\mathbf{j},i+1}^*\psi_{\mathbf{j},i}\rangle$ is the occupation number of mode $\mathbf{j}=(j_x, j_y, j_z)$, averaged over imaginary time, cf.~\eq{occnum}, i.e., summed over $i$, cf.~\App{Matsubara}. 
The factor in \Eq{rho} containing cosines is a Jacobian that takes into account the sine-spacing of the lattice momenta.
%==============================================================
%
\begin{figure*}[t]	
	\includegraphics[width=0.9\columnwidth]{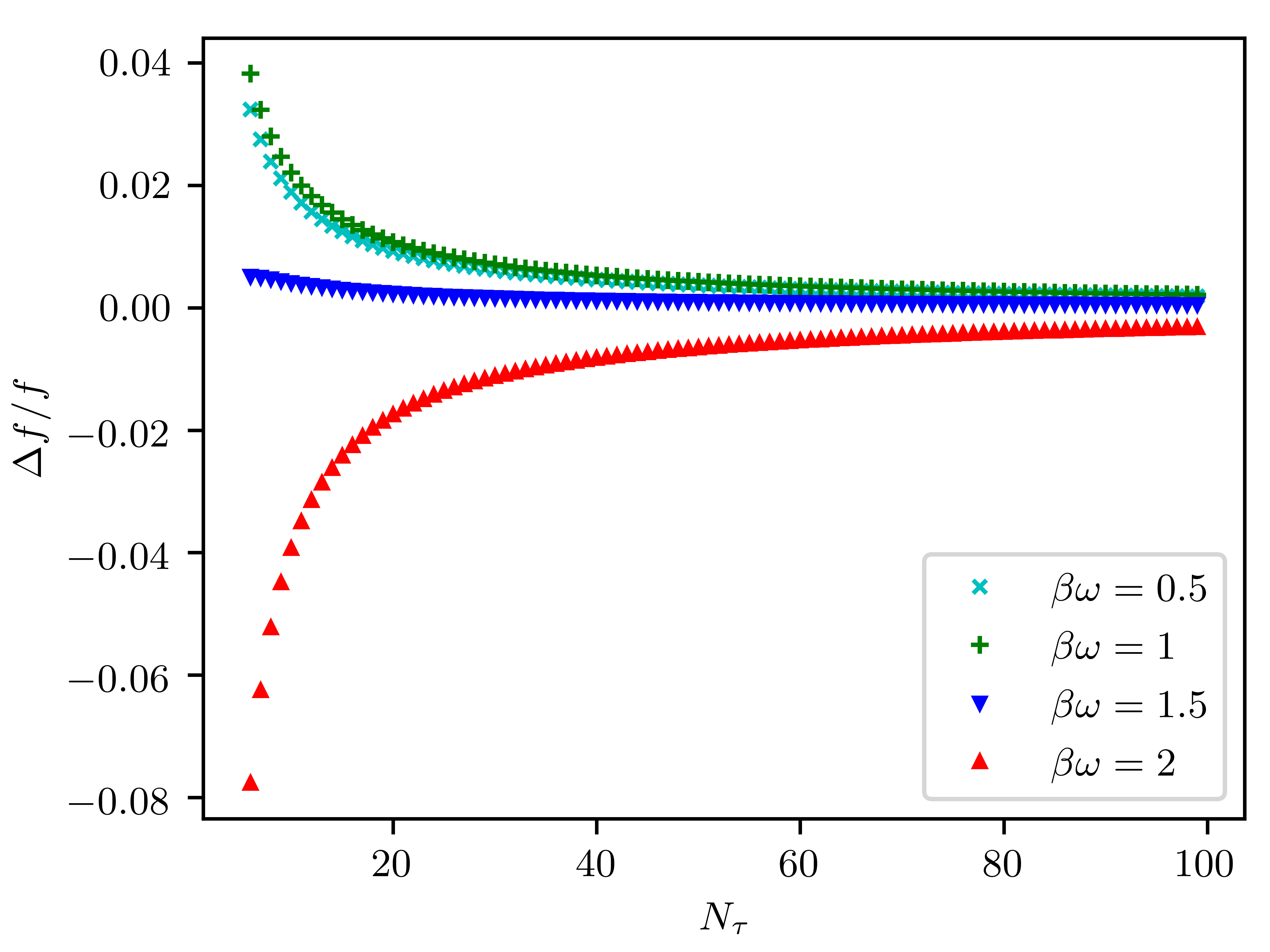}
	\hspace*{0.15\columnwidth}
	\includegraphics[width=0.9\columnwidth]{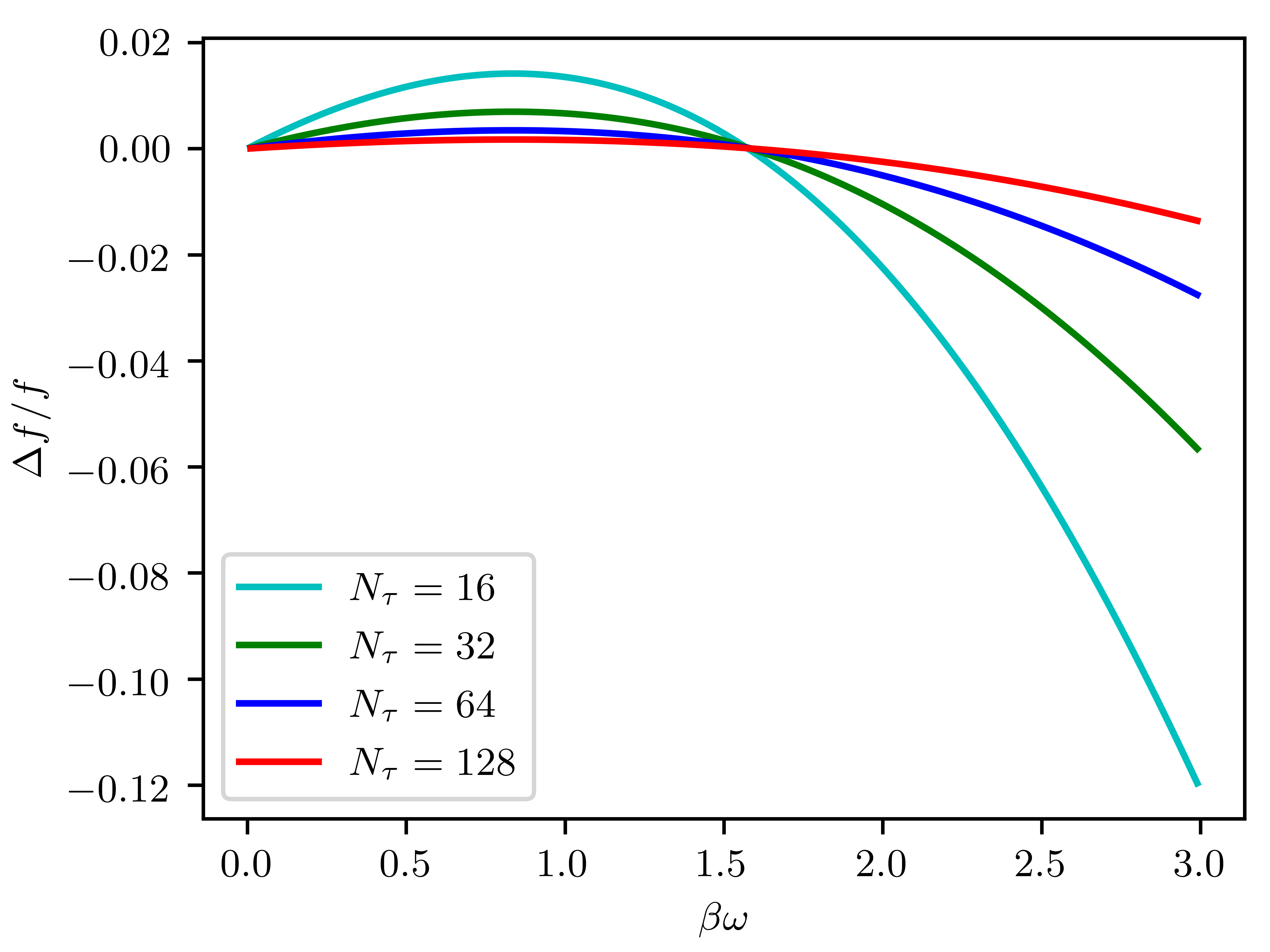}
	\caption{Relative error $\Delta f/f=[f(N_{\tau})-f]/f$ of the particle number $f(N_{\tau})$, \Eq{pi3}, with respect to the continuum value $f$, \Eq{fsinglemode}, (left panel) for different values of $\beta\omega$ in dependence of the truncation order ${N_\tau}$ and (right panel) for different truncation orders ${N_\tau}$ as function of $\beta\omega$, with ${N_\tau}$ increasing from the (on the left side) uppermost curve to the lowermost one.}
	\label{fig:Matsubara}
\end{figure*}
%==============================================================

%======================================================================================
\subsection{\label{app:Matsubara}Truncated Matsubara expansion of particle spectrum and dispersion}
For the purpose of numerical simulations, the imaginary time interval $[0,\beta]$ must be discretized. Here we want to discuss in more detail the errors induced by this discretization. 
Consider, e.g., the expectation value $f(\mathbf{k},\tau)=\langle a^\dagger_{\mathbf{k}}(\tau)a_{\mathbf{k}}(\tau)\rangle$ of the particle number operator in mode $\mathbf{k}$, for a single-component gas. 

In the continuum, dropping any momentum indices, the expectation value can be written as
%$\tau$-average \eq{occnum} 
%
\begin{align}
	\label{eq:fsinglemode}
	f=\langle a^\dagger a\rangle
	=\frac{\text{Tr}\left\{\mathrm{e}^{-\beta\omega a^\dagger a}a^\dagger a\right\}}{\text{Tr}\left\{\mathrm{e}^{-\beta\omega a^\dagger a}\right\}}
	\,.
\end{align}
Evaluating this expression in the number-state basis gives the Bose-Einstein distribution,
\begin{align}
	\langle a^\dagger a\rangle=\frac{1}{\mathrm{e}^{\beta\omega}-1}
	\,.
\end{align}
Within the path integral formulation, the thermal expectation value of the number operator is represented by an imaginary-time path integral, the discretized Feynman representation of which reads
\begin{align}
	\label{eq:pi-number}
	\langle a^\dagger a\rangle
	=\lim\limits_{{N_\tau}\to\infty}\frac{1}{{N_\tau}}\sum_j
	\frac{\int \prod_i d\psi_i^* d\psi_i\,\mathrm{e}^{-S_{N_\tau}[\boldsymbol{\psi}]}\,\psi_{j+1}^*\psi_j}
	{\int \prod_i d\psi_i^* d\psi_i\,\mathrm{e}^{-S_{N_\tau}[\boldsymbol{\psi}]}}
	\,,
\end{align}
with the time-discretized action ($a_{\tau}=\beta/{N_\tau}$)
\begin{align}
	S_{N_\tau}[\boldsymbol{\psi}]
	=\sum_{i=0}^{N_\tau-1}\left(\psi_i^*\psi_i-\psi_{i+1}^*\psi_i+a_{\tau}\omega\, \psi_{i+1}^*\psi_i\right)
	\,,
\end{align}
where the index $N_\tau$ is identified with the index $0$.

\begin{figure}
	\includegraphics[width=0.9\columnwidth]{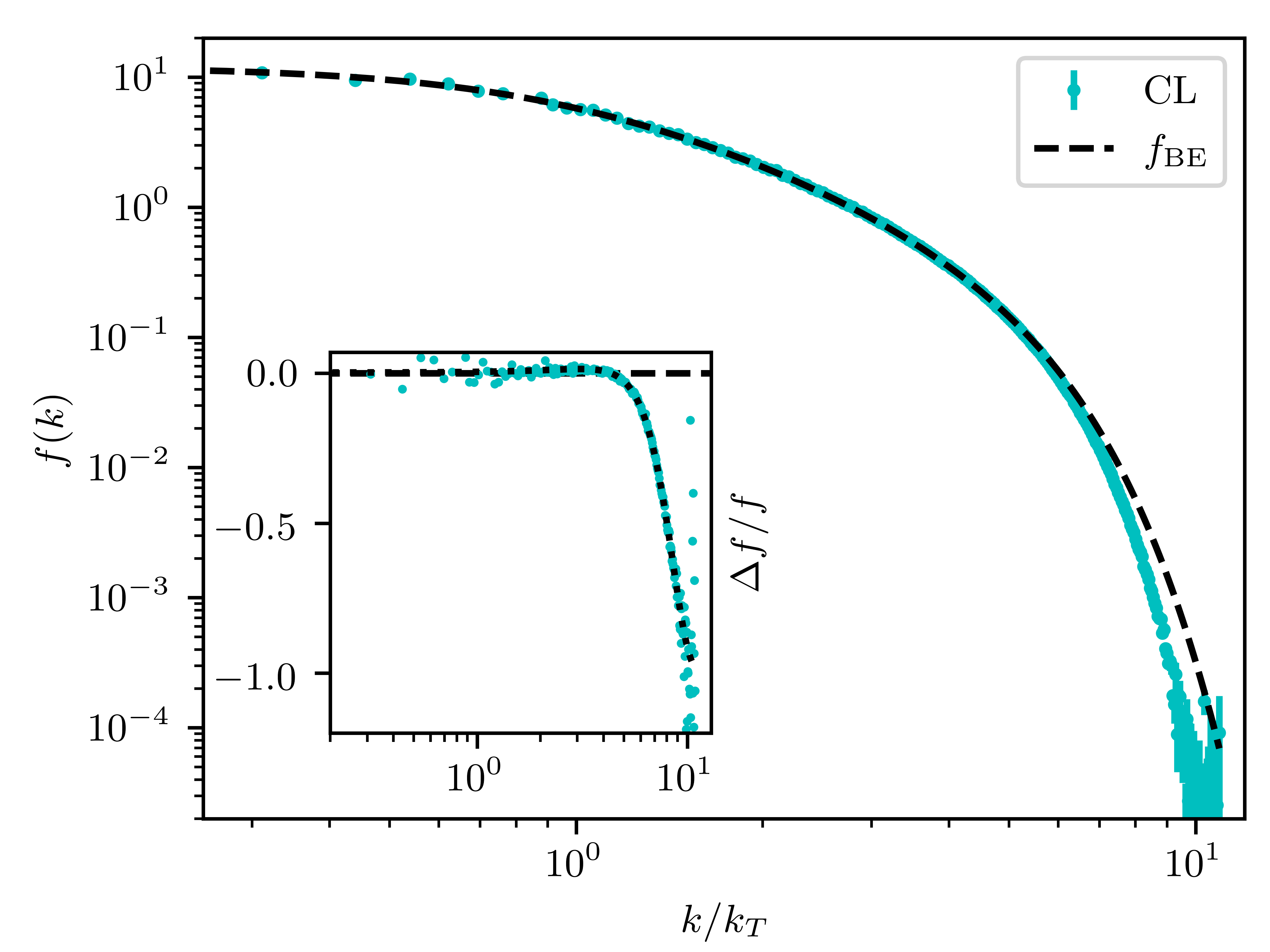}
	\caption{The same angle-averaged momentum spectrum as shown in \Fig{freespectrum}, but including the highest momenta in the ``corners'' of the momentum lattice, i.e. up to $k_\mathrm{max}=2\sqrt{3}\,a_\mathrm{s}^{-1}$. 
	}
	\label{fig:freespectrumwcorners}
\end{figure}

Expanding the fields on the finite time interval $[0,\beta]$ in terms of Matsubara modes,
\begin{align}
	\psi_j=\frac{1}{\sqrt{\beta}}\sum_{n=0}^{N_{\tau}-1} \tilde{\psi}_n\,\mathrm{e}^{\mathrm{i}\omega_n j a_{\tau}}
	\,,
\end{align}
with the Matsubara frequencies $\omega_n={2\pi n}/{\beta}$, $n=0,\dots, {N_\tau}-1$, the path integral \eq{pi-number} becomes
\begin{align}
	\label{eq:pi2}
	\langle a^\dagger a\rangle
	=\lim\limits_{{N_\tau}\to\infty}\frac{1}{\beta}\sum_{m=0}^{N_{\tau}-1}
	\frac{\int \prod_n d\tilde{\psi}_n^* d\tilde{\psi}_n\,\mathrm{e}^{-S_{N_\tau}[\tilde{\boldsymbol{\psi}}]}\,
	\tilde{\psi}_m^*\tilde{\psi}_m\,\mathrm{e}^{-\mathrm{i}\omega_ma_{\tau}}}
	{\int \prod_n d\tilde{\psi}_n^* d\tilde{\psi}_n\,\mathrm{e}^{-S_{N_\tau}[\tilde{\boldsymbol{\psi}}]}}
\end{align}
with 
\begin{align}
	S_{N_\tau}[\tilde{\boldsymbol{\psi}}]
	=\sum_{n=0}^{N_{\tau}-1}\left(\frac{1-\mathrm{e}^{-\mathrm{i}\omega_na_{\tau}}}{a_{\tau}}
	+\omega \mathrm{e}^{-\mathrm{i}\omega_na_{\tau}}\right)\tilde{\psi}_n^*\tilde{\psi}_n
	\,.
\end{align}
Performing the Gaussian integrals yields
\begin{align}
	\label{eq:pi3}
	\langle a^\dagger a\rangle
	&=\lim\limits_{{N_\tau}\to\infty}\frac{1}{\beta}\sum_n
	\frac{\mathrm{e}^{-\mathrm{i}\omega_na_{\tau}}}
	{a_{\tau}^{-1}({1-\mathrm{e}^{-\mathrm{i}\omega_na_{\tau}}})+\omega\, \mathrm{e}^{-\mathrm{i}\omega_na_{\tau}}}
	\nonumber\\
	&=\lim\limits_{{N_\tau}\to\infty}\sum_n\frac{1}{{N_\tau}(\mathrm{e}^{2\pi \mathrm{i} n/{N_\tau}}-1)+\beta\omega }
	\nonumber\\
	&\equiv\lim\limits_{{N_\tau}\to\infty}f(N_\tau)\,.
\end{align}
One shows in a similarly way that the dispersion $\omega$, defined in \Eq{omegataudiscretized}, for finite 
$N_\tau$ becomes
\begin{align}
	\omega(N_\tau)
	=\frac{N_\tau}{\beta}\sqrt{-\frac{1}{f(N_\tau)}\sum_n
	\frac{2\mathrm{e}^{-2\pi \mathrm{i} n/{N_\tau}}-\mathrm{e}^{-4\pi \mathrm{i} n/{N_\tau}}-1}
	{{N_\tau}(\mathrm{e}^{2\pi \mathrm{i} n/{N_\tau}}-1)+\beta\omega }}
	\,.
\end{align}
In the continuum limit, ${N_\tau}\to\infty$, the series \eq{pi3}  converges to $1/(\mathrm{e}^{\beta\omega}-1)$. 
On the lattice, however, the series needs to be truncated at some finite ${N_\tau}$, such that an error occurs which depends on ${N_\tau}$, but also on the value of $\beta\omega$. 
\Fig{Matsubara} shows the numerically computed truncation error for several combinations of ${N_\tau}$ and $\beta\omega$. 
In the right panel, one sees that the relative error for small $\beta\omega$ is positive but turns negative at some point and quickly becomes rather large (while the absolute error still decreases). 
This illustrates the needs in temporal resolution for a precise evaluation of small occupation numbers as well as of the corresponding dispersions, the limits of which can be seen in Figs.~\fig{freespectrum}--\fig{belowtrans_disp}. In fact, for large enough $\beta\omega$, the relative error on the particle number can even become $-100\%$, as can be seen from \Fig{freespectrumwcorners}, which shows a version of \Fig{freespectrum} including the part of the spectrum up to $k_\mathrm{max}=2\sqrt{3}\,a_\mathrm{s}^{-1}$.
%==============================================================
%==============================================================

%======================================================================================
\subsection{Finite-resolution corrections near criticality}
\label{app:ExtrDensity}
On the computational lattice, systematic deviations of observables from their continuum counterparts are caused by both the finite number of Matsubara modes and the finite spatial lattice resolution. 
Both mainly affect the UV properties of the system and therefore, in describing the critical behaviour near the phase transition, are expected to give a small correction only. 
Furthermore, for a weakly interacting gas above the transition, the spectrum is well approximated by an ideal-gas Bose-Einstein distribution with a shifted chemical potential, see \Sect{IntBECabovePT}. 
It is thus reasonable to account for the described systematic errors in the following way: 
For a free gas at $\mu\to 0$ in a box of fixed size $\mathcal{V}=L^{3}$, we numerically determine the deviation $\delta \rho=\rho(N_\mathrm{s}\to\infty,N_{\tau}\to\infty)-\rho(N_\mathrm{s},N_{\tau})$ between the density on the lattice and in the continuum (but for finite system size $\mathcal{V}$). 
$\delta\rho$ is then added to the density determined from numerical data for the interacting system. 
For the parameters as chosen in \Sect{IntBGatCriticality}, i.e. a $64^3\times 16$ lattice at $T=1.25\,a_\mathrm{s}^{-1}$, we obtain $\delta\rho=0.00154\,a_\mathrm{s}^{-3}$. 
This is a rather small correction, but it is significant in determining the constant $c$.

%======================================================================================
%======================================================================================
\section{Bogoliubov theory of the U$(\mathcal{N})$ symmetric Bose gas \label{app:bogol}}
In this appendix, we provide, for conciseness, some details on the Bogoliubov mean-field description of an $\mathcal{N}$-component Bose gas described by the U$(\mathcal{N})$ symmetric Hamiltonian \eq{Hamiltonian}.
For a complete discussion in the context of spinor Bose gases, see, e.g., Refs.~\cite{Kawaguchi2012a.PhyRep.520.253,Stamper-Kurn2013a.RevModPhys.85.1191}.
In momentum space, the ``Kamiltonian'' entering the grand-canonical ensemble, reads
\begin{align}
	\hat{K}\equiv\hat{H}-\mu \hat{N}=
	&\sum_\mathbf{k}\sum_\alpha\left(\frac{\mathbf{k}^2}{2m}-\mu\right)a^\dagger_{\mathbf{k},\alpha}a_{\mathbf{k},\alpha}
	\nonumber\\
	&+\frac{g}{2\mathcal{V}}\sum_{\mathbf{k}\mathbf{k}'\mathbf{q}}\sum_{\alpha\alpha'}
	a^\dagger_{\mathbf{k}+\mathbf{q},\alpha}a^\dagger_{\mathbf{k}'-\mathbf{q},\alpha'}
	a_{\mathbf{k},\alpha}a_{\mathbf{k}',\alpha'}
	\,,
\end{align}
where $\alpha,\alpha'=1,\dots,\mathcal{N}$ enumerate the different field components, $g$ is the coupling and $\mathcal{V}$ the volume. 
Making the Bogoliubov ansatz $a_{\mathbf{k},\alpha}=\sqrt{\mathcal{V}\rho_{0}}\,\delta_{\mathbf{k}0}+a_{\mathbf{k},\alpha}(1-\delta_{\mathbf{k}0})$ and expanding to quadratic order in the non-zero mode operator, this becomes
\begin{align}
	\label{eq:coupled}
	\hat{K}=
	&\left(\frac{\mathcal{N}g}{2}\rho_{0}-\mu\right)\mathcal{N} \mathcal{V}\rho_0
	+\sum_{\mathbf{k}\neq 0}\sum_\alpha\left(\frac{\mathbf{k}^2}{2m}+\mathcal{N} g\rho_0-\mu\right)
	a^\dagger_{\mathbf{k},\alpha}a_{\mathbf{k},\alpha}
	\nonumber\\
	&+g\rho_0\sum_{\mathbf{k}\neq 0}\sum_{\alpha\alpha'}
	\left[a^\dagger_{\mathbf{k},\alpha}a_{\mathbf{k},\alpha'}
	+\frac12\left(a_{\mathbf{k},\alpha}a_{-\mathbf{k},\alpha'}
	+a^\dagger_{\mathbf{k},\alpha}a^\dagger_{-\mathbf{k},\alpha'}\right)
	\right]
	\nonumber\\
	&+\mathcal{O}(a_{\mathbf{k},\alpha}^3)
	\,,
\end{align}
where $\rho_0$ is the condensate density, which is chosen equal in each component, such that the total condensate density is $\mathcal{N}\rho_0$. 
In order to diagonalize in the component degree of freedom, we introduce
\begin{align}
	\label{eq:trafo1}
	B_{\mathbf{k}}
	&=\frac{1}{\sqrt{\mathcal{N}}}\sum_\alpha a_{\mathbf{k},\alpha}
	\\
	\label{eq:trafo2}
	b_{\mathbf{k},\beta}
	&=\sum_\alpha w_\beta^\alpha a_{\mathbf{k},\alpha}
	\qquad \beta=1,\dots,\mathcal{N}-1
	\,,
\end{align}
where the vectors $\mathbf{w}_\beta$ form an arbitrary orthonormal basis of the $\mathcal{N}-1$ dimensional orthogonal complement of the $\mathcal{N}$-component vector $(1,1,\dots,1)^T/\sqrt{\mathcal{N}}$. 
Thereby we obtain
\begin{align}
	\label{eq:decoupled}
	\hat{K}=
	&\left(\frac{\mathcal{N}g}{2}\rho_{0}-\mu\right)\mathcal{N} \mathcal{V}\rho_0
	+\sum_{\mathbf{k}\neq 0}\left(\frac{\mathbf{k}^2}{2m}+2\mathcal{N} g\rho_0-\mu\right)
	B^\dagger_{\mathbf{k}}B_{\mathbf{k}}
	\nonumber\\
	&+\frac{\mathcal{N}g\rho_0}{2}\sum_{\mathbf{k}\neq 0}\left(
	B_{\mathbf{k}}B_{-\mathbf{k}}+B^\dagger_{\mathbf{k}}B^\dagger_{-\mathbf{k}}
	\right)
	\nonumber\\
	&+\sum_{\mathbf{k}\neq 0}\sum_{\beta}\left(\frac{\mathbf{k}^2}{2m}+\mathcal{N} g\rho_0-\mu\right)
	b^\dagger_{\mathbf{k},\beta}b_{\mathbf{k},\beta}
	+\mathcal{O}(a_{\mathbf{k},\alpha}^3)
	\,.
\end{align}
The equivalence of \eq{coupled} and \eq{decoupled} can be easily seen by plugging the transformation \eq{trafo1} and \eq{trafo2} into \eq{decoupled} and exploiting the fact that $(1,1,\dots,1)^T/\sqrt{\mathcal{N}}$ and the $\mathbf{w}_{\beta}$ form an orthonormal set.
In leading-order mean-field approximation the energy is minimized at the chemical potential $\mu=\mathcal{N}g\rho_0$. 
Upon inserting this, \eq{decoupled} becomes
\begin{align}
	\hat{K}=
	&-\frac{g}{2}\mathcal{N}^2\rho_0^2\mathcal{V}
	+\sum_{\mathbf{k}\neq 0}\left(\frac{\mathbf{k}^2}{2m}+\mathcal{N} g\rho_0\right)
	B^\dagger_{\mathbf{k}}B_{\mathbf{k}}
	\nonumber\\
	&+\frac{\mathcal{N}g\rho_0}{2}\sum_{\mathbf{k}\neq 0}\left(
	B_{\mathbf{k}}B_{-\mathbf{k}}+B^\dagger_{\mathbf{k}}B^\dagger_{-\mathbf{k}}
	\right)
	\nonumber\\
	&+\sum_{\mathbf{k}\neq 0}\sum_{\beta}\frac{\mathbf{k}^2}{2m}
	b^\dagger_{\mathbf{k},\beta}b_{\mathbf{k},\beta}
	+\mathcal{O}(a_{\mathbf{k},\alpha}^3)
	\,,
\end{align}
which describes Bogoliubov modes $B_\mathbf{k}$ and $\mathcal{N}-1$ free Goldstone excitations $b_{\mathbf{k},\beta}$. 
The total occupation number $f(\mathbf{k})=\sum_a\langle|\psi_a(\mathbf{k})|^2\rangle$ for $\mathbf{k}\neq 0$ can then be written as 
\begin{align}
	f(\mathbf{k})
	=&\sum_\alpha a^\dagger_{\mathbf{k},\alpha}a_{\mathbf{k},\alpha}
	=B^\dagger_\mathbf{k} B_\mathbf{k}+\sum_\beta b^\dagger_{\mathbf{k},\beta}b_{\mathbf{k},\beta}
	\nonumber\\
	=&\ \frac{1}{2}\left[\frac{\varepsilon(\mathbf{k})+\mathcal{N}g\rho_0}{\omega(\mathbf{k})}-1\right]
	+\frac{\varepsilon(\mathbf{k})+\mathcal{N}g\rho_0}{\omega(\mathbf{k})(\mathrm{e}^{\beta\omega(\mathbf{k})}-1)}
	\nonumber\\
	&\ +\frac{\mathcal{N}-1}{\mathrm{e}^{\beta\varepsilon(\mathbf{k})}-1}
	\,,
\end{align}
with Bogoliubov and free dispersions
\begin{align}
	\omega(\mathbf{k})
	&=\sqrt{\varepsilon(\mathbf{k})\left(\varepsilon(\mathbf{k})+2\mathcal{N}g\rho_0\right)}\\
	\varepsilon(\mathbf{k})
	&=\frac{\mathbf{k}^2}{2m}
	\,.
\end{align}
%

%======================================================================================
%======================================================================================
\section{\label{app:finitesize}Possible finite-size effects in the extraction of $\Delta T_c$}
\begin{figure}[h!]
	\includegraphics[width=0.9\columnwidth]{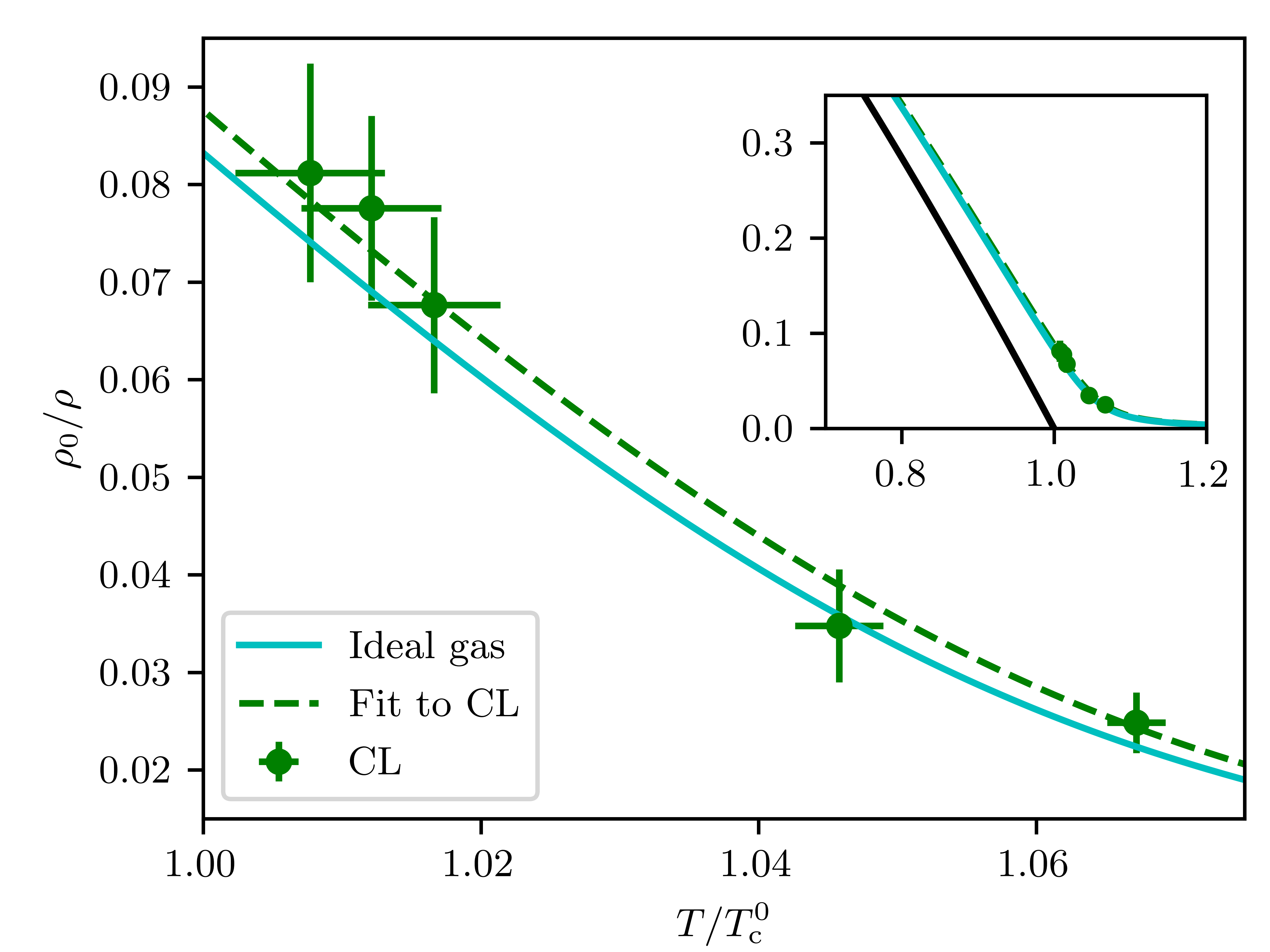}
	\includegraphics[width=0.9\columnwidth]{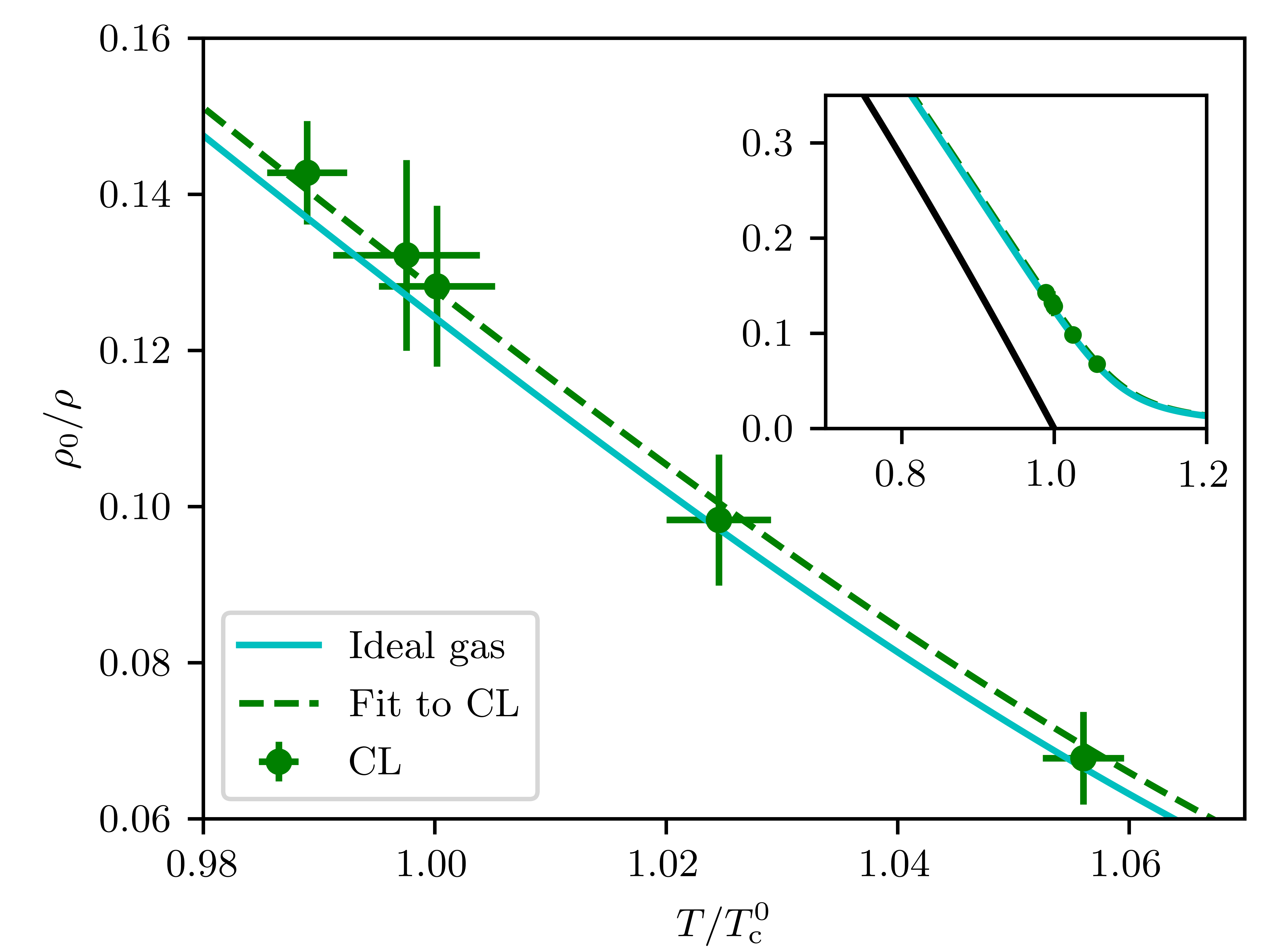}
	\caption{The same as in Fig. \ref{fig:shift}, but on a $48^3$ (upper panel) and $32^3$ (lower panel) lattice. 
	}
	\label{fig:shift3248}
\end{figure}
In our extraction of the transition temperature shift due to interaction effects we refrained from performing a systematic finite-size analysis. Numerical results by \cite{Kashurnikov2001critical} (who performed such a systematic analysis and defined the shift in the finite system in a manner similar to ours) suggest that for the system size that was employed here ($20.2$ thermal wave lengths $\lambda_{T}$), finite size corrections are already significantly smaller than our statistical error. Namely, they find the shift of the critical temperature to scale with system size $L$ as
\begin{align}
\frac{\Delta T_\text{c}}{T_\text{c}^0}\sim \frac{1}{1+b_0 g (2m)^{3/2}T^{1/2}+\frac{a_1+b_1 g(2m)^{3/2}T^{1/2}}{(Lm^2Tg)^{1.038}}}
\end{align} 

with $a_1=1.29$, $b_0=0.123$, $b_1=0.744$ \footnote{Note that, in Ref.~\cite{Kashurnikov2001critical}, constants were chosen  $2m=T=1$ in the factors multiplying $b_{0}$ and $b_{1}$.}.
For our parameters and system size, we obtain from this formula a deviation of $14.6\%$ between the finite-$L$ and the $L\to \infty$ value for the shift, which is only half our statistical error ($27.8\%$), which justifies our procedure. 

Our own simulations corroborate said weak dependence on $L$. 
To demonstrate this, we perform the same analysis as in the main text but for the smaller $48^3$ and $32^3$  lattices, for the exact same chemical potentials. The result is shown in \Fig{shift3248}. Here we obtain for the shift
\begin{align}
\Delta T_c/T_c^0&=0.0036\pm 0.0014\qquad (48^3\,\text{lattice})\\
\Delta T_c/T_c^0&=0.0031\pm 0.0008\qquad (32^3\,\text{lattice})\,.
\end{align}
Within the errors, this agrees with the result for the $64^3$ lattice, suggesting that finite-size effects already play a minor role.      
%==============================================================

%==============================================================

%

\end{document}